\documentclass{article}

\usepackage[preprint]{neurips_2026}

\usepackage[utf8]{inputenc}
\usepackage[T1]{fontenc}
\usepackage[hidelinks]{hyperref}
\usepackage{url}
\usepackage{microtype}
\usepackage{xcolor}
\usepackage{graphicx}
\usepackage{amsmath}
\usepackage{amssymb}
\usepackage{amsthm}
\usepackage{bm}
\usepackage{booktabs}
\usepackage{algorithm}
\usepackage{algpseudocode}
\usepackage{tikz}
\usepackage{soul}
\usepackage[small]{caption}
\usepackage{subcaption}
\usepackage{tabularray}
\usepackage{tabularx}
\usepackage{threeparttable}
\usepackage{adjustbox}
\usepackage{array}
\usepackage{enumerate}
\usepackage{float}

\newtheorem{theorem}{Theorem}
\newtheorem{lemma}[theorem]{Lemma}
\newtheorem{proposition}[theorem]{Proposition}
\theoremstyle{remark}
\newtheorem{remark}[theorem]{Remark}

\colorlet{blue}{black}

\title{Differentiable Combinatorial Auction Design: Optimizing Revenue-Welfare-Fairness Frontiers With Graph Attention Networks}

\author{%
  Mai Pham \\
  Thayer School of Engineering, Dartmouth College \\
  \texttt{mai.p.pham.th@dartmouth.edu}
  \And
  Vikrant Vaze \\
  Thayer School of Engineering, Dartmouth College \\
  \texttt{vikrant.s.vaze@dartmouth.edu}
  \And
  Peter Chin \\
  Thayer School of Engineering, Dartmouth College \\
  \texttt{peter.chin@dartmouth.edu}
}

\begin{document}
\maketitle

\begin{abstract}
    Designing optimal auctions for allocating public infrastructure—such as airport slots and energy resources—is challenging due to interdependent resources and complex side-constraints. While the Vickrey-Clarke-Groves (VCG) auction guarantees maximum welfare, it cannot maximize revenue or accommodate the multi-objective trade-offs inherent in public infrastructure systems. We propose a differentiable, constraint-aware framework for combinatorial auction design that enables end-to-end learning of fractionally feasible, approximately strategy-proof mechanisms.

    Our primary contribution is a combinatorial constraint enforcement technique that ensures fractional feasibility by construction, bypassing the need for penalty-based approximations. We introduce \textit{CAGraph}, a novel graph-based architecture designed to capture dependencies between items and bidders, alongside \textit{CANet} and \textit{CAFormer} as deep learning baselines, for combinatorial settings. We also provide a Matching Pursuit-inspired decomposition that maps fractional allocations to feasible discrete allocations and empirically evaluates the resulting discretization gap.
    
    In synthetic benchmarks and an airport slot-divestiture case study, respectively, the learned mechanisms achieve up to 73.5\% and 180\% higher expected revenue than VCG, under small empirical regret. Finally, we visualize the Pareto frontiers for revenue, welfare, and fairness, providing regulators with a decision-support tool to design infrastructure auctions that balance profitability, efficiency and equity.
\end{abstract}

\section{Introduction}

    Designing optimal auctions has significant real-world implications for the allocation of public infrastructure—such as airport slots, energy markets, and transportation networks—as well as digital assets like online advertising. While mechanism design provides a theoretical framework for achieving desirable outcomes like efficiency and revenue maximization, analytically deriving optimal mechanisms is often intractable in high-dimensional strategic environments. Traditional approaches, such as the Vickrey-Clarke-Groves (VCG) mechanism, offer theoretical guarantees for welfare maximization but are unable to maximize revenue or handle complex side-constraints.

    Recent advances in differentiable economics offer a promising data-driven path forward, utilizing deep neural networks to flexibly approximate optimal auction rules. However, extending these methods to combinatorial auctions—where bidders value bundles of items—remains a formidable challenge due to the complex, non-convex nature of allocation constraints. This paper introduces a differentiable, constraint-aware framework for combinatorial mechanism design that enables the end-to-end learning of fractionally feasible, approximately strategy-proof, and multi-objective mechanisms that balance revenue, efficiency, and fairness.

    \subsection{Background}
    \subsubsection{The Allocative Challenge}
    Combinatorial auctions play an important role in allocating scarce public resources of high economic value, such as spectrum licenses and airport slots. However, designing effective mechanisms for these domains presents two fundamental challenges. First, bidders often exhibit complex valuation structures characterized by strong complementarities and substitutabilities—for instance, an airline values a landing slot only if paired with a corresponding takeoff slot. This necessitates the use of bundle-based valuations, where the package value differs from the sum of its parts. Second, the allocation of public goods is inherently multi-objective. Governments often prioritize allocative efficiency (welfare) to ensure that resources are assigned to their most productive users. However, this focus can lead to significant shortcomings in raising public funds or compensating divesting parties. Furthermore, traditional auction mechanisms often fail to uphold fairness, which is essential to protect disadvantaged agents and prevent monopolistic dominance. A critical gap exists in the literature: traditional mechanisms prioritize single objectives in isolation—VCG for welfare \citep{Vickrey, Clarke, Groves}, Myerson for revenue in single-item settings \citep{Myerson}, or Random Serial Dictatorship (RSD) for fairness \citep{random_serial_1998, BOGOMOLNAIA2001295}. Extending these frameworks to multi-objective settings is analytically intractable and computationally prohibitive, as even the single-objective winner determination problem is NP-hard \citep{sandholm_2003_amd}.

    \subsubsection{Application to Airport Slot Allocation}
    We focus on airport slot allocation, a domain characterized by extreme scarcity and high economic value, with slot portfolios often valued in the millions of dollars. The current administrative regime, governed by the International Air Transport Association (IATA) guidelines and ``Grandfather Rights,'' perpetuates inefficiencies by allowing incumbents to hold slots in perpetuity, often leading to hoarding and suboptimal utilization. While theoretical proposals such as congestion pricing \citep{carlin_and_park_1970, daniel_1995, vaze2012airline} and primary auctions \citep{rassenti_1982, ball_2006} have been suggested, they have failed to gain widespread adoption due to their complexity and the lack of fairness guarantees \citep{jacquillat_vaze_2018, konstantinos2019}, which are essential for political and regulatory acceptability.

    \subsubsection{Methodological Gap}
    A significant methodological gap persists in addressing these multi-faceted challenges. Traditional mechanism design \citep[e.g.,][]{Vickrey, Myerson} excels at optimizing single objectives but struggles to incorporate complex side-constraints like fairness. Conversely, computational approaches for combinatorial auctions must explicitly handle winner-determination and feasibility constraints, which are known to be computationally challenging in high-dimensional bundle spaces \citep{Rothkopf1998, Sandholm_2002, deVriesVohra2003}. While deep learning, specifically Differentiable Economics \citep{regretnet_paper}, offers a promising avenue due to its adaptability to data, standard neural network architectures do not automatically encode the combinatorial structures. This motivates architectures with problem-specific inductive biases, i.e., the architectural assumptions that help the model respect the structures of the problem that are necessary to satisfy the constraints \citep{Battaglia2018, Cappart2023}. 
    
    This paper addresses these challenges by examining and quantifying the trade-offs between three key objectives—Revenue, Welfare, and Fairness. Our work contributes to multi-objective auction design, with real-world case studies focusing on airport resource allocation as an exemplar application domain, by leveraging deep neural networks for constrained resource allocation.

    \subsection{Contributions}
    \begin{enumerate}
        \item \textbf{Data-driven Multi-Objective Optimization.} We provide a data-driven framework to discover high-performing, approximately strategy-proof mechanisms for combinatorial auctions where analytical results are unavailable. This framework models and quantifies the trade-offs between Revenue, Welfare, and Fairness, enabling policymakers to quantify the \textit{Price of Fairness} and calibrate trade-offs between protecting disadvantaged agents and maintaining efficiency.

        \item \textbf{Constraint-Aware Neural Architectures.} We introduce differentiable constraint enforcement techniques for the fractional relaxation of combinatorial allocation constraints. This enables us to extend existing architectures to the combinatorial setting (CANet, CAFormer) and propose \textbf{CAGraph}, a novel graph-based architecture that generalizes the winner determination problem to a learned, differentiable set-packing task.

        \item \textbf{Resolution-Aware Implementation.}
        We identify the implementability gap created by fractional combinatorial allocations and propose a Matching Pursuit-inspired decomposition that constructs an implementable lottery over feasible deterministic schedules. The method is used as an approximation procedure, with ex-post discretization effects evaluated empirically.

        \item \textbf{Empirical Evaluation on a Real-World Case Study.} We evaluate CAGraph on an airport slot-allocation case study using real-world datasets, showing tunable revenue–welfare–fairness trade-offs relative to VCG and Random Serial Dictatorship. In the Washington National Airport (DCA) slot-divestiture setting, the conditional full-assignment implementation variant reaches mean realized revenue of \$536.3k, a 284.2\% increase relative to VCG.

    \end{enumerate}

    \subsection{Paper Outline}
    The remainder of the paper is organized as follows. Section 2 reviews related literature. Section 3 formalizes the multi-objective combinatorial auction problem. Section 4 presents the differentiable combinatorial auction framework, including constraint enforcement, neural architectures, and the discrete implementation step. Section 5 reports computational results, while Section 6 presents a real-world airport slot allocation case study. Section 7 concludes with directions for future work.
    
    \section{Related Literature}
    Our research relates to multi-objective auction theory, differentiable economics, and airport resource allocation. In each subsection, we summarize the relevant literature in one of these streams.

    \subsection{Multi-Objective Auction Theory}
    Classical auction theory has primarily focused on single-objective optimization. The Vickrey-Clarke-Groves (VCG) mechanism serves as the standard solution for guaranteeing dominant-strategy incentive compatibility (DSIC) and exact welfare maximization \citep{Vickrey, Clarke, Groves}. In contrast, Myerson's optimal auction provides the theoretical foundation for revenue maximization in single-item settings based on virtual valuations \citep{Myerson}.

    However, even in bilateral trade, efficiency can conflict with incentive and participation requirements. The Myerson-Satterthwaite Theorem establishes that no mechanism can simultaneously achieve ex-post efficiency, Bayesian incentive compatibility, interim individual rationality, and budget balance \citep{MyersonSatterthwaite83}. This inherent tension is further exemplified by the Revenue-Welfare Non-Coincidence: even in symmetric settings with uniform valuations, the revenue-maximizing allocation rule (which may withhold items to drive up competition) differs from the welfare-maximizing rule (which always allocates to the highest value) \citep{Myerson}.

    Beyond efficiency and revenue, fairness is increasingly recognized as a critical requirement in public resource allocation. Unlike private markets, public domains often require an equitable distribution of utility to ensure political and regulatory acceptability. Fairness can be quantified by metrics such as the Jain Index, Gini Coefficient, and Coefficient of Variation. Recent works have explored fairness-welfare trade-offs \citep[e.g.,][]{konstantinos2019} and welfare-revenue trade-offs via Automated Mechanism Design \citep{sandholm_2003_amd}. However, most existing multi-objective approaches are restricted to single-item or simple unit-demand settings and do not extend to the combinatorial constraints inherent in resource allocation domains, such as airport slot allocation.

    In contrast to this literature, we do not seek an analytical characterization of a single optimal mechanism. Instead, we provide a data-driven framework for empirically tracing revenue--welfare--fairness frontiers in combinatorial auctions with side constraints. This allows the designer to evaluate policy-relevant trade-offs even when closed-form optimal mechanisms are unavailable.

    \subsection{Airport Resource Allocation}
    The allocation of airport slots is currently governed by administrative procedures such as the IATA Worldwide Slot Guidelines (WSG). A central feature of this regime is ``Grandfather Rights,'' which allow incumbent airlines to retain slots in perpetuity, provided they utilize them sufficiently. While providing operational stability, this system creates significant barriers to entry and encourages slot hoarding, impeding allocative efficiency \citep{ball_2006}. Theoretical alternatives, including congestion pricing \citep{carlin_and_park_1970, daniel_1995, vaze2012airline} and combinatorial auctions \citep{rassenti_1982, ball2018}, have been proposed to address these rigidities but have struggled to gain political traction due to implementation complexities and equity concerns.
    
    In practice, regulatory interventions must balance conflicting objectives that pure market mechanisms often ignore. Regulators aim to maximize allocative efficiency (welfare) while frequently needing to generate public funds (revenue) or ensure competitive equity (fairness) \citep{jacquillat_vaze_2018, konstantinos2019}. This multi-objective complexity was exemplified by the 2011 slot transfers at Washington National Airport (\textit{DCA}) and LaGuardia Airport (\textit{LGA}), where regulators sought to foster competition \citep{DOT2011SlotWaiver, FederalRegister2011SlotWaiver} by requiring divestitures that expanded access for new-entrants and limited-incumbent carriers. More recently, Spirit Airlines’ 2026 bankruptcy led to a competitive auction of 22 slots at LGA, with JetBlue named the successful bidder at \$58.5 million\citep{FederalRegister2026SpiritSlots,JetBlue2026Q2}. Consequently, acceptable auction mechanisms must balance revenue generation against allocative efficiency and bidder surplus so that the resulting allocation remains politically feasible and commercially viable for participating airlines.

    We model airport slot divestiture as a constrained combinatorial auction and use real-world airline data to quantify how alternative learned mechanisms trade off revenue, allocative efficiency, and fairness. Rather than prescribing a single regulatory rule, the framework provides a decision-support tool for comparing policies.
    
    \subsection{Deep Neural Networks for Mechanism Design}
    Differentiable Economics has emerged as a promising paradigm for automated mechanism design, leveraging deep learning to learn optimal mechanism rules directly from data. This approach, exemplified by RegretNet \citep{regretnet_paper}, parameterizes the mechanism as a neural network and optimizes an objective function—such as revenue maximization—via gradient descent, subject to regret penalties that encourage approximate dominant-strategy incentive compatibility (DSIC). Concurrently, Graph Neural Networks (GNNs) have shown great promise in solving combinatorial optimization problems \citep{cappart2023combinatorial} and winner determination problems in energy markets \citep{ahmed2023combinatorial}. The differentiable mechanism-design framework pioneered by RegretNet has demonstrated remarkable success in recovering known optimal auctions in single-item settings and discovering new mechanisms where analytical solutions are unknown.

    However, extending this paradigm to high-stakes domains like airport slot allocation presents significant combinatorial challenges. Bidders value specific combinations of items (e.g., paired landing and takeoff slots), creating an exponential valuation space. Crucially, allocations must satisfy strict integrality and feasibility constraints: each item can be allocated at most once, and agents often demand valid bundles rather than fractional fragments. Standard feed-forward neural networks or standard transformer architectures often fail to produce outputs that strictly adhere to these complex feasibility constraints, frequently outputting impractical fractional assignments.

    This creates a critical need for specialized neural architectures that enforce fractional feasibility by design, and for a separate discrete implementation step.  Birkhoff’s theorem provides exact lottery implementability for simple assignment structures, whereas overlapping combinatorial constraints need not admit such a decomposition \citep{budish2013designing}, we therefore measure and report the residual of a greedy discrete approximation.
    
    Our work bridges this gap by introducing graph-based architectures and differentiable resolution layers that inherently satisfy these combinatorial feasibility constraints.

    \section{Problem Statement and Model}
    \label{sec:CA_design}

    \paragraph{General auction setting.} We study a setting with one seller, a set $N$ of $n$ bidders, and a set $M$ of $m$ items for sale. An auction rule is a pair: an allocation rule $g(.)$ and a payment rule $p(.)$. In the differentiable mechanism studied here, $g_{iS}(b)\in[0,1]$ denotes the probability, or fractional allocation weight, that bidder $i$ receives bundle $S$ under bid profile $b$, and $p_i(b)$ denotes bidder $i$'s payment under bid profile $b$. Deterministic allocations are treated as a case in which $g_{iS}(b)\in\{0,1\}$.


\paragraph{Non-additive valuations.}
Let $K=\{S\subseteq M:S\neq \emptyset\}$ denote the set of all non-empty bundles, with $|K|=k=2^m-1$. Each bidder $i$ has a private valuation function $v_i:K\rightarrow \mathbb{R}_{\ge 0}$, where $v_i(S)$ is bidder $i$'s value for bundle $S$. 
Because $K$ is finite, we represent this function by the vector
$
v_i = \bigl(v_{iS}\bigr)_{S\in K} \in \mathbb{R}_{+}^{k},
$
where $v_{iS}\equiv v_i(S)$. The valuation profile is
$
v=(v_1^T,\ldots,v_n^T)\in \mathbb{R}_{\ge 0}^{n\times k},
$
and is private information unknown to the auctioneer. Let $V_i\subseteq \mathbb{R}_{\ge 0}^{k}$ denote bidder $i$'s valuation space. We assume $v_i$ is drawn from distribution $F_i$ over $V_i$. The auction environment is \textit{symmetric} if $F_i=F$ for all $i\in N$, and \textit{asymmetric} if bidders' valuation distributions differ.

\paragraph{Bids and quasi-linear utilities.}
Each bidder reports a bid vector $b_i=(b_{iS})_{S\in K}\in \mathbb{R}_{\ge 0}^{k}$, where $b_{iS}$ is bidder $i$'s reported bid for bundle $S$. Let $b=(b_1^T,\ldots,b_n^T)\in \mathbb{R}_{\ge 0}^{n\times k}$ denote the bid profile. We assume quasi-linear utilities in expectation over the mechanism's stochastic allocation rule. Given allocation probabilities $g_i(b)=(g_{iS}(b))_{S\in K}$ and payment $p_i(b)$, bidder $i$'s utility is
$
u_i(v_i;b)=\sum_{S\in K} g_{iS}(b)v_{iS}-p_i(b).
$
For a deterministic allocation in which bidder $i$ receives bundle $S$, this reduces to $u_i(v_i;b)=v_{iS}-p_i(b)$.

    \paragraph{Incentive Compatibility and Regret.} 
    A mechanism is a \textit{dominant-strategy incentive compatible (DSIC)} mechanism if each bidder's utility is maximized when reporting their true valuation, regardless of what other bidders report.
\begin{equation}
    \label{dsic}
    \begin{aligned}
    u_i(v_i;(v_i,b_{-i}))
    &\ge u_i(v_i;(b_i,b_{-i})),\\[-1mm]
    &\forall v_i,b_i\in V_i,\quad
      \forall b_{-i}\in V_{-i},\\[-1mm]
    &\forall i\in N.
    \end{aligned}
\end{equation}

Following \citet{duetting2015payment}, we quantify the violation of DSIC using \textit{expected ex-post regret}, defined for each bidder as follows.
\begin{equation}
    \begin{aligned}
    \operatorname{rgt}_i
    =\mathbb{E}_{v\sim F}\!\bigl[&
    \max_{b_i\in V_i}u_i(v_i;(b_i,v_{-i}))\\[-1mm]
    &-u_i(v_i;(v_i,v_{-i}))\bigr],\\[-1mm]
    &\forall i\in N.
    \end{aligned}
\end{equation}


\paragraph{Individual Rationality (IR).} A mechanism is \textit{ex-post individually rational} if each bidder receives a nonnegative utility when participating truthfully. 
\begin{equation}
    u_i (v_i; (v_i, b_{-i})) \ge 0, \: \forall v_i\in V_i,\ \forall b_{-i}\in V_{-i}, \: \forall i \in N
    \label{ir}
\end{equation}

The incentive audit below applies to the learned fractional mechanism and does not automatically transfer through the discrete-resolution and conditional-pricing steps. Section~\ref{subsec:practical_implementation} therefore audits the decomposition residual and realized economic outcomes, but does not claim that post-resolution regret has been established.
\paragraph{Combinatorial Feasibility.}

We first state the deterministic feasibility constraints and then relax them to obtain the stochastic/fractional feasibility region used by the differentiable mechanism. Let $z_{iS}$ specify whether a bundle \( S \) is allocated to bidder \( i \) (\( z_{iS} = 1 \)) or not (\( z_{iS} = 0 \)). The standard CA formulation requires the following constraints to hold.
\begin{equation}
    \sum_{i\in N} \sum_{S \in K: j\in S} z_{iS} \leq 1 \quad \forall j \in M
    \label{item_sum}
\end{equation}
\begin{equation}
    \sum_{S \in K} z_{iS} \leq 1 \quad \forall i \in N
    \label{agent_sum}
\end{equation}
\begin{equation}
    z_{iS} \in \{0, 1\} \quad \forall i \in N, S \in K
    \label{binary}
\end{equation}

Constraints \eqref{item_sum} ensure that each item \( j \) is allocated to at most one combination of a bidder and a bundle, implicitly guaranteeing that no two bidders can receive overlapping bundles. 
Constraints \eqref{agent_sum} ensure that each bidder \( i \) is assigned at most one bundle.
Any allocation satisfying these constraints is considered \textit{combinatorially feasible}.

To make the solution space differentiable, we relax the binary constraints \eqref{binary}. 

\begin{equation}
    z_{iS} \in [0, 1] \quad \forall i \in N, S \in K
    \label{real}
\end{equation}
Fractional feasibility is enforced during neural-network training. Because it does not necessarily imply exact implementability as a lottery over deterministic combinatorial allocations, Section~\ref{sec:method} discusses the resulting implementation gap and the discrete decomposition procedure.

\paragraph{Multi-Objective Formulation}
The auctioneer maximizes a specific design goal (e.g., revenue, welfare, or a combination) while respecting the constraints of the mechanism. For a valuation profile $v$ and a fractional allocation $Z=g(v)$, social welfare is given as follows.
\[
\operatorname{Wel}(v,Z)=\langle v,Z\rangle
=\sum_{i\in N}\sum_{S\in K}v_{iS}z_{iS}.
\] 

For a given valuation profile \( v \), let $\mathcal{O}(v, g, p)$ denote the auxiliary objective function to be maximized. The mechanism observes reports \(b\), not true valuations \(v\). For training and evaluation, the expected design objective is computed on truthful valuation profiles by setting \(b=v\). The multi-objective mechanism design problem can be formulated as an optimization of the allocation and payment rules as follows.
\[
\max_{\textbf{g}, \textbf{p}} \quad \mathbb{E}_{v \sim F} [\mathcal{O}(v, g(v), p(v))]
\]
subject to \eqref{dsic}, \eqref{ir}, \eqref{item_sum}, \eqref{agent_sum}, \eqref{real}.

Fairness can be incorporated as a soft training target. We measure fairness over bidder utilities using the Jain index,
\(J(u)=\frac{(\sum_{i\in N}u_i)^2}{n\sum_{i\in N}u_i^2}\),
where larger values indicate a more equal distribution of utility, and \(J=1\) corresponds to equal utility across bidders. We use the Jain index because it provides a scale-invariant measure of the concentration of bidder surplus. In the fairness experiments, the implemented equality-style penalty targets a selected value of \(J(u)\).

Following the RegretNet framework \citep{regretnet_paper}, we model optimal auction design as a learning task in which the allocation and payment rules are parameterized by neural networks, allowing us to address multi-bidder, multi-item scenarios where analytical solutions are unknown.

\section{Solution approach} \label{sec:method}
Throughout this section, all operations are applied profile-wise, and batch dimensions are suppressed unless explicitly stated. The bid profile is denoted by $b\in\mathbb{R}_{\ge0}^{n\times k}$, the fractional agent-bundle allocation matrix by $Z=(z_{iS})\in[0,1]^{n\times k}$, and the payment vector by $p\in\mathbb{R}_{\ge0}^{n}$. The neural architectures produce unnormalized allocation scores, which are transformed by the constraint-enforcement layer into feasible fractional allocations. We use superscripts such as ``agent,'' ``bundle,'' and ``agent-bundle'' to distinguish the intermediate arrays used in this transformation.

\begin{figure}[t]
    \centering
    \includegraphics[width=\textwidth]{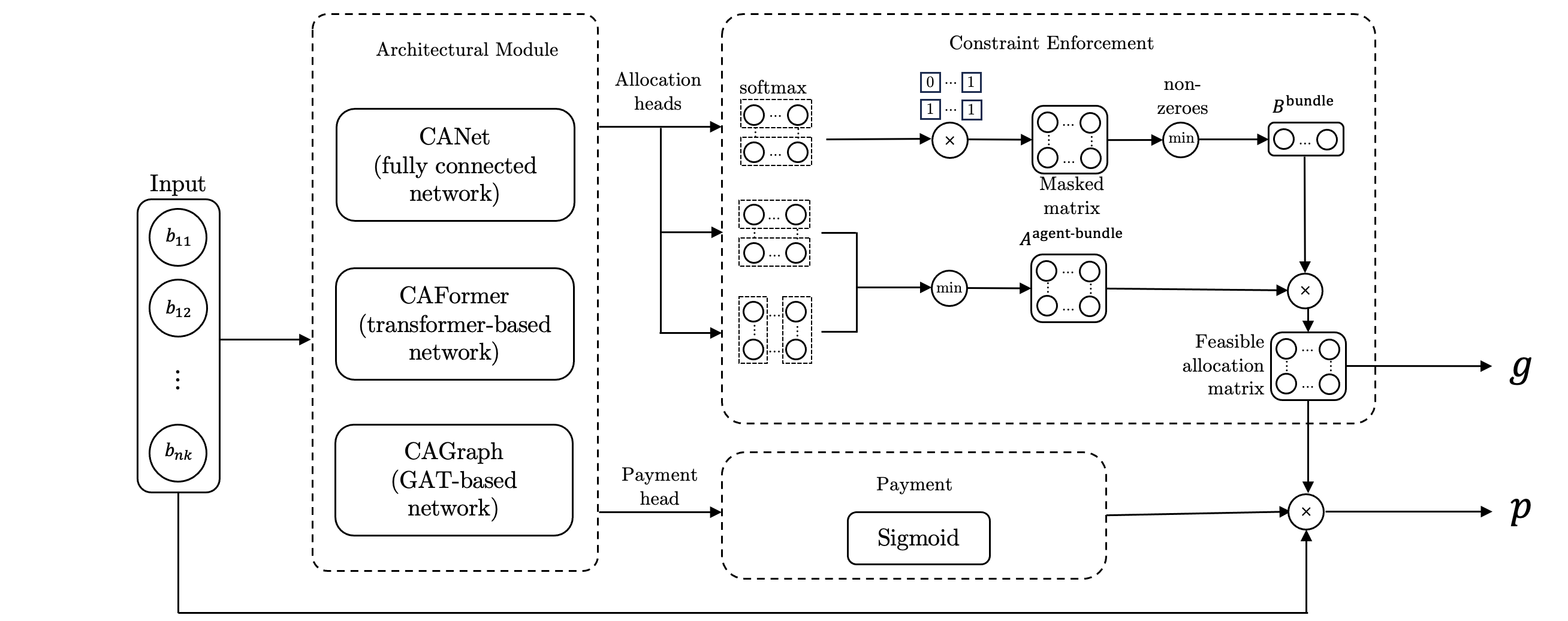}
    \caption{Overview of the proposed differentiable auction framework. The architecture enforces fractional feasibility via constraint factorization and optimizes revenue–welfare–fairness trade-offs.}
    \label{fig:unified}
\end{figure}

\subsection{Constraint Enforcement via Factorization}
\label{subsec:constraint_enforcement}
Recall from Section~\ref{sec:CA_design} that the allocation must satisfy constraints
\eqref{item_sum}, \eqref{agent_sum}, and \eqref{real}. We decompose the allocation matrix \(Z \in [0,1]^{n \times k} \) into the product of two matrices. Let $B\in\mathbb{R}^{m\times k}$ and $A, A'\in\mathbb{R}^{n\times k}$ denote unnormalized score matrices produced by the allocation network for the item--bundle and agent-bundle normalization steps, respectively. The constraint-enforcement layer takes unnormalized network outputs and maps them to a fractional allocation matrix $Z\in[0,1]^{n\times k}$. Let
$B^{\mathrm{bundle}}\in[0,1]^{n\times k}$ be the bundle-availability factor, and let $A^{\mathrm{agent-bundle}}\in[0,1]^{n\times k}$ denote the agent-bundle assignment factor (defined in
\eqref{eq:bundle_factor}--\eqref{eq:agent_bundle_factor}). The final allocation is obtained by the element-wise product
\begin{equation}
    Z
    =
    B^{\mathrm{bundle}}
    \odot
    A^{\mathrm{agent-bundle}}.
    \label{eq:factorized_allocation}
\end{equation}

The item-to-bundle allocation matrix \( B^{\text{bundle}} \) is derived from the output of the Allocation Network, normalized to represent item-wise probabilities. To construct \( B^{\text{bundle}} \), we define the incidence matrix \( I \in \mathbb{R}^{m \times k} \), where $I_{jS} = 1$ if item $j$ is included in bundle $S$, and 0 otherwise. 
The initial bundle allocation matrix \( B \in \mathbb{R}^{m \times k} \) is adjusted by mapping bundles to items using a softmax function across the \(k\) bundles for each item.
\begin{equation}B^{\text{adjusted}} = \text{softmax}_K B \odot \mathbf{I}. \label{eq:bundle_factor}
\end{equation}
Here $\odot$ denotes element-wise multiplication. To exclude entries corresponding to items not contained in a bundle, zero values in \( B^{\text{adjusted}} \) are replaced with a large constant \( C_{\text{mask}} \). For each bundle \(S\), we compute the minimum value across its constituent items \(B^{\min}_S=\min_{j\in S}B^{\text{adjusted}}_{jS}\) and replicate \( B^{\text{min}} \) to include the agent dimension to get \(B^{\mathrm{bundle}}_{iS}=B^{\min}_S\) for all \(i\in N\).
    

Let $A,A'\in\mathbb{R}^{n\times k}$ denote two agent-bundle score matrices produced by
the allocation network. We normalize $A$ across agents and $A'$ across bundles as follows.
\begin{equation}
A^{\text{agent}} = \text{softmax}_N (A), \quad A^{\text{bundle}} = \text{softmax}_K(A')
\label{eq:agent_bundle_softmax}
\end{equation}
The agent-bundle assignment factor is then obtained via an element-wise minimum as follows.
\begin{equation}
    A^{\mathrm{agent-bundle}}
    =
    \min\{A^{\mathrm{agent}},A^{\mathrm{bundle}}\}.
    \label{eq:agent_bundle_factor}
\end{equation}

    The allocation matrix \( Z \) is then given by \eqref{eq:factorized_allocation}. We now formally establish that this structural decomposition restricts the network's output space to feasible fractional allocations.
    
    \begin{theorem}[Guaranteed Combinatorial Feasibility]
    \label{thm:feasibility}
    For any input bid profile $b$, the allocation matrix $\textbf{Z}$ produced by the factorized constraint layer satisfies the following conditions.
    \begin{enumerate}
        \item \textbf{Item-wise Feasibility:} $\sum_{i \in N} \sum_{\substack{S\in K:\\ j\in S}} z_{iS} \le 1, \quad \forall j \in M$
        \item \textbf{Agent-wise Feasibility:} $\sum_{S \in K} z_{iS} \le 1, \quad \forall i \in N$
    \end{enumerate}
    \end{theorem}
    
    See Appendix~\ref{app:feasibility_proof} for the complete proof.
    \subsection{Enhancing Existing Architectures} \label{subsec:existing_archs}

\subsubsection{CANet Architecture} \label{subsubsec:canet}
CANet is a deep feed-forward multilayer perceptron (MLP) that maps the flattened valuation profile $b \in \mathbb{R}^{nk}$ to allocation and payment decisions. The network consists of an input linear projection to a hidden dimension $H$, followed by $L$ fully connected layers with LeakyReLU activations. To support multi-objective optimization, the architecture splits into two task-specific heads: (i) an \textit{Allocation Head} that outputs unnormalized logits, which are processed by the constraint factorization layer (Section \ref{subsec:constraint_enforcement}) to ensure feasibility; and (ii) a \textit{Payment Head} that projects the hidden representation to a fractional vector $\tilde{p} \in [0, 1]^n$ via Sigmoid activation.
    
\subsubsection{CAFormer Architecture} \label{subsubsec:caformer}
To enhance expressivity and ensure permutation-equivariance, we introduce CAFormer, a Transformer-based architecture that processes the auction state as two parallel streams of exchangeable embeddings: an \textit{Item Stream} $\mathbf{X}^{\text{item}} \in \mathbb{R}^{q \times m \times H}$ and an \textit{Agent Stream} $\mathbf{X}^{\text{agent}} \in \mathbb{R}^{q \times n \times H}$, where $q$ is the batch size and $H$ is the hidden embedding dimension. These streams pass through $L$ layers of Multi-Head Self-Attention (MHSA) and a specialized \textit{Cross-Attention} mechanism. Unlike \textit{RegretFormer} \citep{regretformer_paper}, our architecture uses agent representations to query item representations, allowing the model to learn complex demand structures before producing agent–bundle logits for the constraint enforcement layer. The final output is processed via an invariant attention head to ensure the mechanism's results are independent of bidder ordering.
    
\subsection{Proposed New Architecture: CAGraph} \label{subsec:cagraph}
We model the combinatorial auction as a \textit{Conflict Graph} $G = (V, E)$, where each node $v_{iS}$ is an agent-bundle pair $(i, S)$. An edge exists between $(i, S)$ and $(j, S')$ if either $i=j$ or the bundles overlap ($S \cap S' \neq \emptyset$), ensuring exclusivity. This set-packing formulation allows the network to reason about feasibility as a maximum weighted independent set problem. The initial node representation $h^{(0)}_{iS}$ is constructed by concatenating the scalar bid $b_{iS}$ with learnable bidder and bundle embeddings. These features propagate through Graph Attention (GAT) layers \citep{velickovic2018graph} as follows.
\begin{equation}
    h^{(l+1)}_{iS} = \sigma \left( \sum_{(j,S')\in \mathcal{N}(i,S)}
    \alpha^{(l)}_{(i,S),(j,S')}
    \mathbf{W}^{(l)} h^{(l)}_{jS'}
    \right).
    \label{eq:cagraph_gat}
\end{equation}
Here $\mathcal{N}(i,S)\subseteq N\times K$ denotes the set of agent-bundle nodes $(j,S')$ that conflict with node $(i,S)$, and $\alpha^{(l)}_{(i,S),(j,S')}$ is the learned attention coefficient from neighbor $(j,S')$ to node $(i,S)$ at layer $l$. By attending to contested slots (i.e., those included in the high-contention nodes), the network learns a dense representation of market friction. The GAT layers encode agent-bundle conflicts, while feasibility is enforced by the constraint factorization rather than by graph attention alone.
    
\subsection{Training Procedure}
\label{sec:training}
The same training procedure is used for CANet, CAFormer, and CAGraph. For mechanism parameters $\theta$, let $g_\theta$ and $p_\theta$ denote the learned allocation and payment rules. The mechanism observes reports $b$, not true valuations $v$; in the training objective, $g_\theta(v)$ and $p_\theta(v)$ denote the evaluation of the bid-dependent rules at truthful reports $b=v$. Let $\operatorname{Obj}(\theta)=\mathbb{E}_{v\sim F}[\mathcal{O}(v,g_\theta(v),p_\theta(v))]$. The implemented objective uses a log-transformed economic loss together with adaptive penalty terms:
\begin{equation}
    \min_{\theta}
    \left\{-\log\!\left(1+\operatorname{Obj}(\theta)\right)
    +\sum_{c\in\mathcal{C}}w_cG_c(\theta)\right\},
    \label{eq:training_loss}
\end{equation}
Here $\mathcal{C}$ is the set of penalized design requirements, $G_c(\theta)$ is the empirical extent of the violation of requirement $c$, and $w_c\ge 0$ is its adaptive penalty weight. For example, the regret term penalizes violations of approximate DSIC, while an equality-style fairness penalty is included when a Jain-index target is selected. The weights $w_c$ are updated during training according to the adaptive controller described in Appendix~\ref{app:training}; these penalty weights are unrelated to the GAT weight matrices $W^{(l)}$ in Section~\ref{subsec:cagraph}. The revenue--welfare trade-off used in our multi-objective experiments is summarized as follows.
    
\begin{remark}[Revenue--welfare weighting]
Define $\mathrm{Rev}_\theta(v)=\sum_{i\in N}p_{\theta,i}(v)$ and
$\mathrm{Wel}_\theta(v)=\sum_{i\in N}\sum_{S\in K}v_{iS}g_{\theta,iS}(v)$. For $\lambda \in [0,1]$, define
\[
\operatorname{Obj}_{\lambda}(\theta)
=\mathbb{E}_{v\sim F}\!\left[
\lambda\,\mathrm{Rev}_{\theta}(v)
+(1-\lambda)\,\mathrm{Wel}_{\theta}(v)\right].
\]
Training minimizes
\[
-\log\!\left(1+\operatorname{Obj}_{\lambda}(\theta)\right)
+\sum_{c\in\mathcal C}w_cG_c(\theta).
\]
Hence, $\lambda$ traces an empirical revenue--welfare frontier. 
The resulting changes in allocation reflect the trade-off between revenue and welfare; we characterize this trade-off empirically rather than claiming a closed-form solution in the combinatorial, approximately DSIC setting.
\end{remark}

\subsection{Implementability and Discrete Decomposition}
\label{subsec:implementability}

The neural architectures described above output fractional allocation matrices
$Z \in [0,1]^{n \times k}$, whereas airport slots are indivisible and must
ultimately be implemented as deterministic schedules. Thus, fractional feasibility is not
sufficient; the learned allocation must either be representable as a lottery over feasible
integer allocations or approximated by such a lottery.

In simple assignment settings, this implementation gap is closed by the Birkhoff--von Neumann theorem. After augmenting the problem with dummy agents or items so that a doubly sub-stochastic assignment can be completed to a doubly-stochastic matrix, every fractional doubly-stochastic matrix can be written as a convex combination of permutation matrices and is therefore implementable as a lottery over deterministic assignments. Combinatorial slot allocation is more difficult because overlapping bundles induce constraints that are not captured by the standard assignment polytope. \citet{budish2013designing} show that universal implementation depends on the structure of the constraints: roughly, constraints must be nested or disjoint rather than partially overlapping. They call such constraint systems \emph{bihierarchies}. When this structure fails, some feasible fractional allocations may not be implementable exactly as lotteries over deterministic feasible allocations. In particular, their necessity result implies the following.
\begin{lemma}[\citealt{budish2013designing}]
If a canonical two-sided constraint structure is not a bihierarchy, then it is not universally implementable.
\label{theobudish}
\end{lemma}

We formalize the connection between this implementability limitation and our slot-allocation setting in Appendix~\ref{app:implementability}. This limitation is relevant here because overlapping slot bundles can induce non-bihierarchical constraint structures. Let $\mathcal{B}$ denote the set of feasible deterministic schedules satisfying the binary constraints \eqref{item_sum}--\eqref{binary}. Then the exact implementation of a fractional allocation $Z$ requires $Z\in \operatorname{conv}(\mathcal{B})$. Fractional feasibility under \eqref{item_sum}, \eqref{agent_sum}, and \eqref{real} does not guarantee this condition.

\begin{proposition}[Implementability limitation]
\label{prop:implementability_limitation}
Fractional feasibility under constraints \eqref{item_sum}, \eqref{agent_sum}, and \eqref{real} does not, in general, imply exact implementability as a lottery over feasible deterministic bundle allocations. In particular, overlapping slot bundles can induce non-bihierarchical constraint structures, for which some feasible fractional points lie outside $\operatorname{conv}(\mathcal{B})$, where $\mathcal{B}$ denotes the set of feasible
deterministic schedules.
\end{proposition}

This motivates an explicit resolution step. Rather than assuming exact implementability,
we construct an implementable lottery over feasible deterministic schedules using a Matching Pursuit-inspired
decomposition. 

Let $R_0=Z$. At iteration $t$, define
\begin{align}
D_t
&\in \arg\max_{D\in\mathcal{B}}
\langle R_t,D\rangle, \label{eq:mp_select}\\
\gamma_t
&= \min_{(i,S)\in \Omega_t} R_{t,iS}, \label{eq:mp_weight}\\
R_{t+1}
&= R_t-\gamma_t D_t, \label{eq:mp_update}
\end{align}
where $\Omega_t=\{(i,S):D_{t,iS}=1\}$.

Because feasibility is downward closed, selected bidder--bundle pairs with zero residual can be removed from a maximizing schedule without changing its objective or violating feasibility. We use this reduced maximizer and terminate if it is empty or if $\gamma_t$ is below the numerical tolerance.
The extracted support after $T$ iterations is $\Lambda_T=\{D_t:\gamma_t>0,\;t=0,\ldots,T-1\}$, where $\gamma_t$ is the raw extraction weight associated with schedule $D_t$; the valid lottery weights are defined below. Here
$\langle R_t,D\rangle=\sum_{i\in N}\sum_{S\in K}R_{t,iS}D_{iS}$
denotes the Frobenius inner product.

\begin{proposition}[Residual reduction]
\label{prop:residual_reduction}
If $R_t\ge 0$ and $\gamma_t$ is chosen by \eqref{eq:mp_weight}, then $R_{t+1}\ge 0$ and
\[
\|R_{t+1}\|_1 \le \|R_t\|_1,
\]
with strict decrease whenever $\gamma_t>0$ and $\Omega_t\ne\emptyset$.
\end{proposition}

See Appendix~\ref{app:residual_reduction_proof} for the proof.

For a valuation profile $v$, the welfare induced by allocation $Z$ is $\langle v,Z\rangle=\sum_{i\in N}\sum_{S\in K}v_{iS}z_{iS}$. After $T$ iterations, set $c_T=\max\{1,\sum_{t=0}^{T-1}\gamma_t\}$ and use lottery weights $\pi_t=\gamma_t/c_T$, assigning the remaining mass to the feasible null schedule. The expected allocation of the implementable lottery is $\widehat Z_T=\sum_{t=0}^{T-1}\pi_tD_t$. This normalization does not require $\sum_t\gamma_t\leq1$ and leaves the extracted weights unchanged when their sum is at most one. Let $E_T=Z-\widehat Z_T$ denote the implementation residual. Suppose $0\le v_{iS}\le v_{\max}$ for all $i\in N$ and $S\in K$.
\begin{equation}
\big|\langle v,Z\rangle-\langle v,\widehat Z_T\rangle\big|
=|\langle v,E_T\rangle|
\le v_{\max}\|E_T\|_1.
\label{eq:welfare_gap}
\end{equation}
The inequality follows directly from H\"older's inequality:
$|\langle v,E_T\rangle|\le \|v\|_\infty\|E_T\|_1\le v_{\max}\|E_T\|_1$.

The bound shows that the economic loss from implementation is proportional to the implementation residual. In particular, if bundle values are bounded by $v_{\max}$, then replacing the fractional allocation $Z$ with the expected allocation $\widehat Z_T$ of the implementable lottery changes expected welfare by at most $v_{\max}\|E_T\|_1$. Thus, $E_T$ provides an interpretable certificate of implementation error: a smaller implementation residual implies a tighter worst-case welfare bound.

The greedy decomposition is not guaranteed to recover the sparsest lottery representation. Rather, it provides a computationally tractable nonnegative sparse approximation: each iteration selects the feasible deterministic allocation most aligned with the current residual and extracts the largest coefficient that preserves nonnegative raw residual mass. Every schedule in the resulting support is feasible and interpretable. The raw residual $R_T$ measures unextracted mass, whereas $E_T$ measures the error of the normalized lottery that is actually implemented.

\paragraph{Scope of guarantees.}
Our guarantees concern fractional feasibility, raw-residual reduction, and an economic bound in terms of the implementation residual. We do not claim universal exact lottery implementability for non-bihierarchical combinatorial environments. When $Z\notin\operatorname{conv}(\mathcal{B})$, the decomposition is an approximation procedure, and its ex-post revenue, welfare, and fairness effects are evaluated empirically. Incentive properties of the fractional mechanism do not automatically transfer through decomposition and conditional pricing; post-resolution regret must be audited separately when making incentive claims about the implemented lottery.

\section{Computational Results: Synthetic Benchmarks}\label{sec:computation_results}

We validate our framework on standard distributions to assess its ability to recover optimal solutions and scale to complex bundle valuations. Models are trained for 100,000 iterations across two settings: (i) \textbf{Non-combinatorial} with additive item values drawn from $U[0,1]$, and (ii) \textbf{Combinatorial} with a complementarity term $c_{iS} \sim U[-1,1]$ added to the sum of item values to test non-additive preferences, allowing both super-additive and sub-additive bundle values. For brevity, we denote the non-combinatorial additive setting by (A), the symmetric combinatorial setting by (B), and the asymmetric combinatorial setting by (C). More precisely, let $v^{\mathrm{item}}_{ij}$ be bidder $i$'s value for item $j$ and construct $v_{iS}=\sum_{j\in S}v^{\mathrm{item}}_{ij}+c_{iS}$, where $c_{iS}\sim U[-1,1]$. In Setting B, $v^{\mathrm{item}}_{ij}\sim U[1,2]$ for every bidder and item. In the two-bidder Setting C, $v^{\mathrm{item}}_{1j}\sim U[1,2]$ and $v^{\mathrm{item}}_{2j}\sim U[1,5]$. All draws are independent.
    
\paragraph{Revenue and Welfare Analysis.}
Table 1 summarizes performance in additive settings. While the Myerson baseline for two separate items yields $\sim$0.833 revenue, optimal mechanisms can reach 0.86--0.88 \citep{manelli2006bundling}. \textbf{CANet} and \textbf{CAFormer} match these baselines, confirming that our architectures recover standard auction outcomes. Notably, in the $2 \times 2$ setting, \textbf{CAGraph} achieves revenue of 1.111 , approximately $83\%$ of the expected efficient-welfare benchmark ($\mathbb{E}[\text{Welfare}] \approx 1.33$). The result shows substantial revenue under small measured regret (approximately $0.008$), even though the empirical regret is not reliably interpreted as a proof of DSIC \citep{anshelevich2004price}.
    
\paragraph{Combinatorial Advantages.}
Table 2 reports results for $2 \times 2$ to $2 \times 5$ combinatorial settings. The classical benchmarks include the Affine Maximizer Auction (AMA), Virtual Valuations Combinatorial Auction (VVCA), basic local optimization of AMA (BLAMA), allocation-boosting AMA (ABAMA), and bidder-bundle-boosting VVCA (BBBVVCA) \citep{Sandholm_2015}. While grid-search methods (AMA and VVCA) and \textit{RegretNet} are limited by dimensionality, CAGraph performs well across the tested dimensions. In the complex $2 \times 5$ asymmetric case, \textbf{CAGraph} achieves 12.123 revenue—a 73.5\% improvement over VCG (6.987) and 6.4\% over \textbf{CAFormer}. These results suggest that approximate DSIC can yield substantial revenue gains in the tested combinatorial settings. CAGraph’s conflict-graph representation improves revenue relative to CANet and CAFormer while preserving feasibility.

\begin{table}[t]
\centering
\small
\setlength{\tabcolsep}{7pt}
\renewcommand{\arraystretch}{1.1}
\begin{tabular}{lccc}
\toprule
& \textbf{$2{\times}2$ (A)} & \textbf{$2{\times}3$ (A)} & \textbf{$2{\times}5$ (A)} \\
\midrule
VCG          & 0.667 / 0        & 1.000 / 0        & 1.671 / 0        \\
AMA          & 0.860 / 0        & -- / --          & -- / --          \\
VVCA         & 0.866 / 0        & -- / --          & -- / --          \\
BLAMA        & 0.786 / 0        & 1.222 / 0        & 2.231 / 0        \\
ABAMA        & 0.786 / 0        & 1.255 / 0        & 2.251 / 0        \\
BBBVVCA      & 0.776 / 0        & 1.238 / 0        & 2.242 / 0        \\
RegretNet    & 0.878 / 1e$-$3   & 1.317 / 1e$-$3   & 2.339 / 1e$-$3   \\
RegretFormer & 0.908 / 1e$-$3   & 1.416 / 1e$-$3   & 2.453 / 1e$-$3   \\
CANet        & 0.879 / 1e$-$3   & 1.317 / 1e$-$3   & 2.282 / 4e$-$3   \\
CAFormer     & 0.891 / 1e$-$3   & 1.326 / 1e$-$3   & 2.329 / 5e$-$3   \\
CAGraph      & \textbf{1.111} / 8e$-$3 & \textbf{1.640} / 14e$-$3 & \textbf{2.671} / 10e$-$3 \\
\bottomrule
\end{tabular}
\vspace{1.5mm}
\caption{Non-combinatorial results (per cell: \emph{Revenue} / \emph{Regret}): CANet and CAFormer achieve revenue comparable to RegretNet and RegretFormer, while CAGraph achieves higher revenue.}
\label{non-combinatorial_results}
\end{table}


\begin{table}[t]
\centering
\small
\setlength{\tabcolsep}{4pt}
\renewcommand{\arraystretch}{1.1}
\begin{tabular}{l c c c c c c}
\toprule
& \multicolumn{3}{c}{\textbf{Symmetric (B)}} & \multicolumn{3}{c}{\textbf{Asymmetric (C)}} \\
\cmidrule(lr){2-4}\cmidrule(lr){5-7}
\textbf{Method} & $2\!\times\!2$ & $2\!\times\!3$ & $2\!\times\!5$ & $2\!\times\!2$ & $2\!\times\!3$ & $2\!\times\!5$ \\
\midrule
VCG        & 2.405 / 0     & 3.537 / 0     & 5.838 / 0     & 2.847 / 0     & 4.239 / 0     & 6.987 / 0     \\
AMA        & 2.760 / 0     & — / —          & — / —          & 4.240 / 0     & — / —          & — / —          \\
VVCA       & 2.770 / 0     & — / —          & — / —          & 4.240 / 0     & — / —          & — / —          \\
BLAMA      & 2.630 / 0     & 4.125 / 0     & 7.280 / 0     & 4.080 / 0     & 4.812 / 0     & 8.164 / 0     \\
ABAMA      & 2.630 / 0     & 4.166 / 0     & 7.145 / 0     & 4.010 / 0     & 5.916 / 0     & 11.140 / 0    \\
BBBVVCA    & 2.620 / 0     & 4.105 / 0     & 7.156 / 0     & 4.010 / 0     & 5.898 / 0     & 11.135 / 0    \\
RegretNet  & 2.871 / 1e-3  & — / —          & — / —          & 4.270 / 1e-3  & — / —          & — / —          \\
CANet      & 2.904 / 1e-3  & 4.369 / 2e-3  & 7.304 / 7e-3  & 4.285 / 1e-3  & 6.556 / 1e-3  & 11.393 / 2e-3 \\
CAFormer   & 2.919 / 1e-3  & 4.388 / 2e-3  & 7.318 / 3e-3  & 4.403 / 2e-3  & 6.693 / 1e-3  & 11.534 / 4e-3 \\
CAGraph    & \textbf{3.202} / 1e-3 & \textbf{4.710} / 1e-3 & \textbf{7.774} / 5e-3 & \textbf{4.844} / 3e-3 & \textbf{7.204} / 3e-3 & \textbf{12.123} / 6e-3 \\
\bottomrule
\end{tabular}
\vspace{1.5mm}
\caption{Combinatorial results (per cell: \emph{Revenue} / \emph{Regret}). Grid search methods (AMA and VVCA) become computationally infeasible in higher dimensions. RegretNet results are limited to \( 2 \times 2 \) cases. CANet and CAFormer beat the benchmarks in all settings. But CAGraph consistently outperforms the others in revenue.  In symmetric (B), the revenue upper bound attains 3.548, 5.467, 9.243 for $2 \times 2$ , $2 \times 3$ , $2 \times 5$; in asymmetric (C): 6.184, 9.296, 15.573.}
\label{combinatorial_results}
\end{table}

\section{Practical Applications: Airport Slot Allocation}
\label{sec:case_study}

Capacity-constrained "Level 3" airports (IATA) \citep{iata_wsg} currently rely on administrative "Grandfather Rights," where incumbents retain perpetual tenure via the "80/20 rule." This system incentivizes incumbents to "babysit" underutilized slots to block competition, stifling allocative efficiency. A key regulatory response was the 2011 DOJ-mandated slot divestiture at New York LaGuardia (LGA) and Washington National (DCA) \citep{doj_2011}. Anchoring on this scenario, we model a combinatorial auction among four carriers---JetBlue (B6), Frontier (F9), Spirit (NK), and Sun Country (SY), four candidate slot times (08:00, 10:00, 17:00, and 22:00), which generate six two-slot bundles and one four-slot bundle, at DCA, a realistic regulatory approach targeted at facilitating market entry.
\subsection{Data-Driven Problem Formulation}
\label{subsec:data_formulation}


\subsubsection{Experimental Setup} 


Routes and fares are derived from the Airline Origin and Destination Survey (DB1B) dataset. Seat capacities are obtained from the T-100 Domestic Segment (U.S. Carriers) dataset. All data are downloaded from \url{https://www.transtats.bts.gov}. A profit-maximizing assignment of flights to available slots determines the bundles' valuations.

\paragraph{Parameter Specification.} To instantiate the valuation model, we incorporate specific industry parameters. Revenue is estimated as the product of the average fare, the number of seats, and the average load factor. Operational costs follow the aircraft cost specification of \citet{swan2006aircraft} as a function of distance and seat capacity. For training, we use the calibrated per-mile cost coefficient $c_{\mathrm{mile}}=0.019$; uncertainty in this coefficient is examined out of sample in Section~\ref{subsec:sensitivity}. Profile heterogeneity enters through independently perturbed time-of-day multipliers.
Time-of-day preferences are incorporated to reflect empirically observed variation in airline departure-time demand \citep{lurkin2016continuous}.

\paragraph{Optimization Model.} We model the assignment of directional flights to slots, subject to feasibility and flow balance. The problem is formulated as follows.

\textbf{Given:}
\begin{itemize}
  \item $F$: set of feasible flights for the airline (each with a direction)
  \item $T$: set of time slots in the slot bundle
  \item $V(f, t)$: profit from assigning flight $f \in F$ to slot $t \in T$
  \item $\operatorname{dir}(f) \in \{+1, -1\}$: direction of flight $f \in F$ ($+1$ = departure, $-1$ = arrival)
\end{itemize}

\textbf{Decision Variables:}
\begin{itemize}
  \item $z_{ft} \in \{0,1\}$: equals 1 if flight $f \in F$ is assigned to slot $t \in T$
  \item $w_f \in \{0,1\}$: equals 1 if flight $f \in F$ is not operated (canceled)
\end{itemize}

\textbf{Objective:}
\begin{equation}
\max_{z, w} \quad \sum_{f \in F} \sum_{t \in T} V(f, t) \cdot z_{ft}
\label{eq:val_obj}
\end{equation}

\textbf{Subject to:}
\begin{align}
\sum_{f \in F} z_{ft} &\leq 1, \quad \forall t \in T \label{eq:slot_cap}\\
\sum_{t \in T} z_{ft} + w_f &= 1, \quad \forall f \in F \label{eq:flight_assign}\\
\sum_{f \in F} \sum_{t \in T} \operatorname{dir}(f) \cdot z_{ft} &= 0 \label{eq:flow_balance}
\end{align}

    
\subsubsection{Statistical Properties of Valuations\label{subsubsec:valuation_properties}}
\begin{figure}[t]
    \centering
    \includegraphics[width=\textwidth]{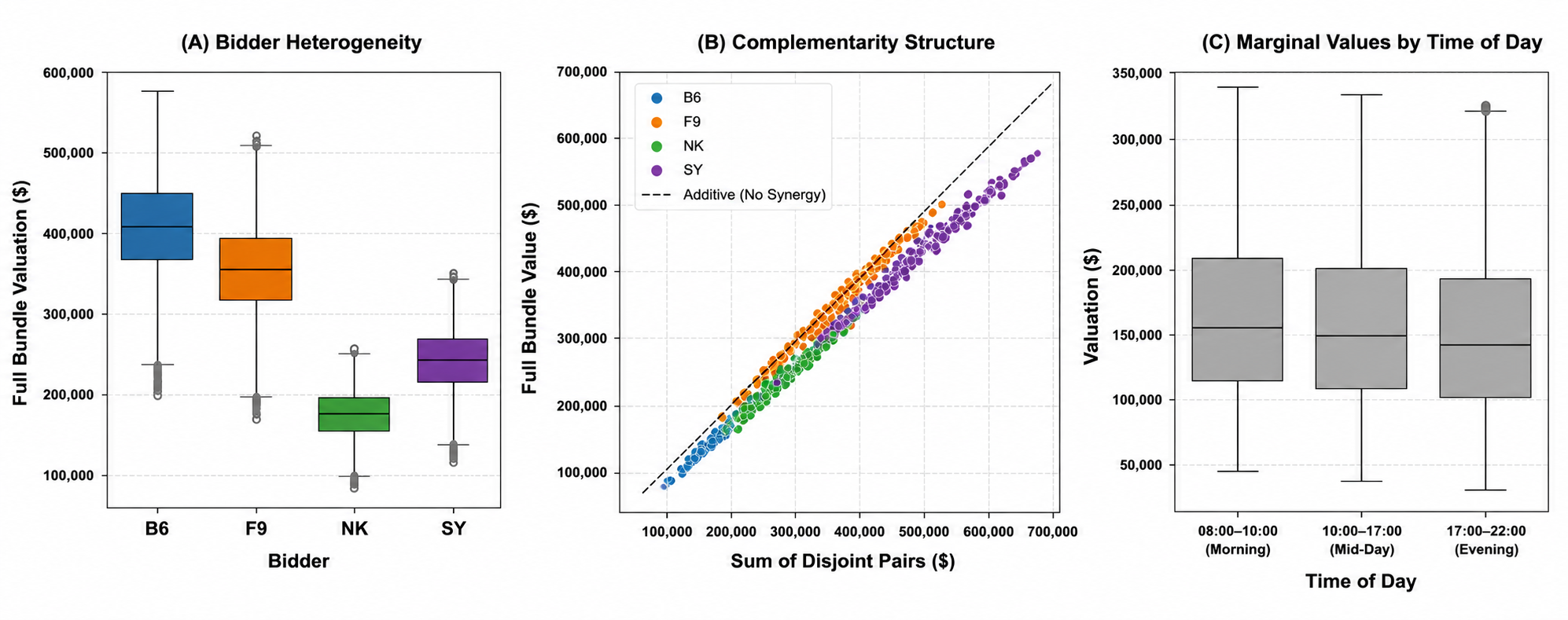}
    \caption{Statistical summary of generated valuations. \textbf{(A)} Bidder asymmetry in full-bundle distributions. \textbf{(B)} The four-slot bundle value is compared with the sum of two disjoint two-slot bundle values; observations below the dashed additive benchmark indicate subadditivity and diminishing returns. \textbf{(C)} Consistency in marginal values across time blocks driven by high-demand anchor slots.}\label{fig:valuation_props}
\end{figure}

Inferred valuations exhibit three critical properties: (i) a right-skewed distribution dominated by high-value peak connections; (ii) systematic subadditivity, with the value of the four-slot bundle generally below the sum of the values of two disjoint two-slot bundles; and (iii) bidder asymmetry, with incumbents (B6, F9) showing higher valuations than niche entrants (NK, SY). This asymmetry, illustrated in Figure~\ref{fig:valuation_props}, justifies our multi-objective approach, as standard VCG risks revenue collapse in such settings. We generate 100,000 profiles with a fixed load factor of 0.7 while sampling the per-mile operating-cost coefficient and time-of-day preference multipliers: 80,000 profiles are used for training and 20,000 for testing.

\subsection{Benchmark Mechanisms}
\label{subsec:benchmarks}
Our framework is compared against two  opposite ends of the welfare-fairness spectrum.

\paragraph{Vickrey-Clarke-Groves (VCG).} As the canonical welfare-maximizing mechanism, \textit{VCG} selects the maximum weighted independent set on the conflict graph. However, it is poorly suited for slot divestitures due to: (i) \textbf{Low revenue}, as VCG payments can be small relative to revenue-maximizing alternatives, providing insufficient compensation; and (ii) \textbf{Market Concentration}, as welfare maximization can concentrate allocations among high-value bidders.

\paragraph{Random Serial Dictatorship (RSD / Lottery).}
The FAA procedure under 14 CFR \S 93.225 uses random draws to establish priority for sequential slot selection. It provides procedural symmetry and does not solicit valuation reports. Standard serial dictatorship is strategy-proof under its direct-choice model, but the administrative priority lottery is not the same as a direct-revelation combinatorial mechanism. Because it ignores monetary valuations, it can sacrifice welfare. For instance, a bidder with a \$10M peak-hour valuation may lose to a \$1M bidder based solely on draw order, failing to allocate capacity to its highest-value use.

\paragraph{Learned Mechanism (CAGraph).} 
Our CAGraph architecture serves as a tunable alternative to static benchmarks, via the mixing parameter $\lambda$, with fairness controlled separately through a soft Jain-index target. It enables regulators to navigate trade-offs among revenue, welfare, and fairness while maintaining approximate incentive compatibility.

To ensure a consistent comparison, Sections~\ref{subsec:empiricalI}-\ref{subsec:revenue_fairness} report \emph{ex-ante} metrics derived from the fractional allocations $Z \in [0,1]^{n \times k}$, with realized ex-post outcomes analyzed in Section~\ref{subsec:practical_implementation}.

\subsection{Empirical Results I: Economic Efficiency and Revenue}
\label{subsec:empiricalI}
\label{subsec:econ_results}

\begin{figure}[t]
    \centering
    \includegraphics[width=\textwidth]{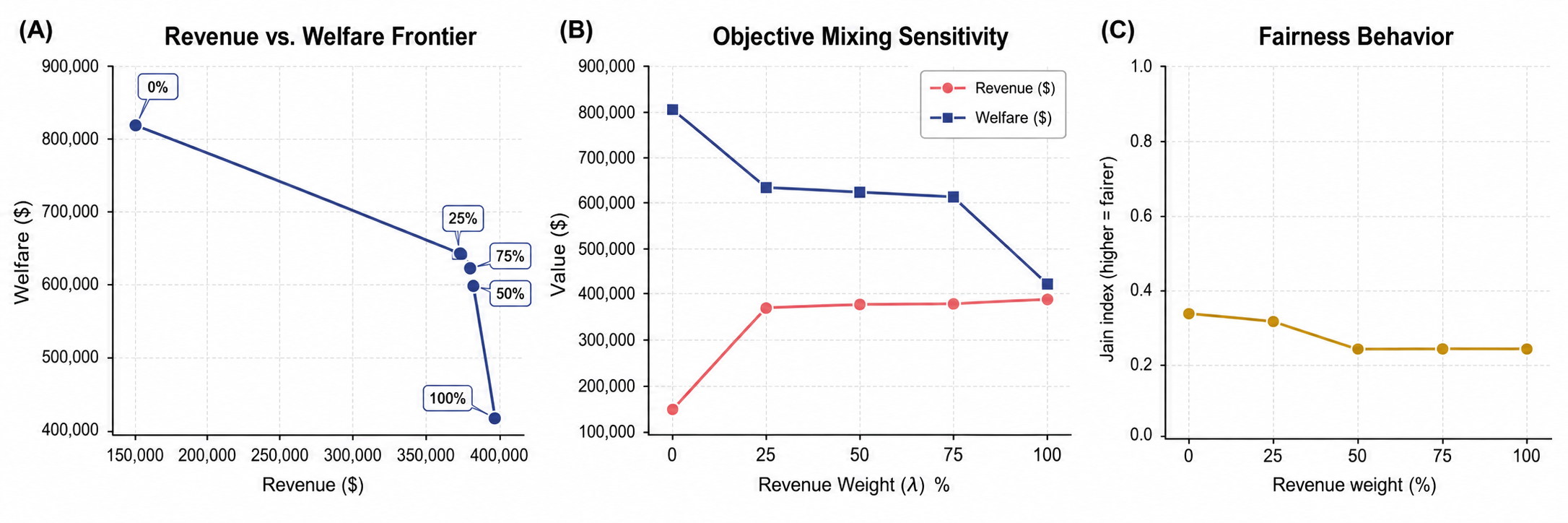}
    \caption{Economic frontiers and sensitivity. \textbf{(A)} Pareto trade-off between revenue and welfare. \textbf{(B--C)} Impact of mixing weight $\lambda$ on economic objectives and the resulting Jain fairness index.}
    \label{fig:rev_wel}
\end{figure}

Using RSD and VCG as baselines, we evaluate how the learned mechanism interpolates between welfare and revenue objectives while maintaining low empirical regret; detailed regret values are reported in Appendix~\ref{app:regret_audit}. Table~\ref{tab:airport_results} and Figure~\ref{fig:rev_wel} show a clear revenue--welfare frontier. At $\lambda=0$, the mechanism nearly matches VCG welfare (\$814k versus \$819k) while generating slightly higher revenue (\$150k versus \$139k). As $\lambda$ increases, revenue rises substantially: at the balanced setting $\lambda=0.5$, revenue reaches \$376k, approximately 171\% higher than VCG, while retaining 77\% of VCG welfare; at $\lambda=1$, revenue reaches \$391k, approximately 181\% higher than VCG, but welfare falls by 49\%. \textcolor{blue}{The frontier exhibits diminishing returns beyond $\lambda=0.25$, which captures approximately 92\% of the maximum observed revenue gain over VCG while retaining 78\% of VCG welfare.} Moderate revenue weighting ($\lambda \leq 0.5$) preserves most welfare, whereas aggressive revenue weighting induces sharper welfare losses. Fairness also declines as $\lambda \rightarrow 1$, with the Jain index falling to 0.25, consistent with revenue maximization concentrating allocations among high-value bidders.

\begin{table}[t]
\centering
\setlength{\tabcolsep}{5pt}
\renewcommand{\arraystretch}{1.1}
\begin{tabular}{l|rr|rrrrr}
\toprule
\textbf{Metric} & \textbf{RSD} & \textbf{VCG} & \boldmath$\lambda{=}0$ & \boldmath$\lambda{=}0.25$ & \boldmath$\lambda{=}0.5$ & \boldmath$\lambda{=}0.75$ & \boldmath$\lambda{=}1$ \\
\midrule
Revenue (\$k) & 0 & 139 & 150 & 370 & 376 & 377 & 391 \\
Welfare (\$k) & 518 & 819 & 814 & 635 & 625 & 615 & 418 \\
Fairness      & 1.00 & 0.25 & 0.35 & 0.32 & 0.25 & 0.25 & 0.25 \\
\bottomrule
\end{tabular}
\vspace{1mm}
\caption{Airport slot auction results. The learned mechanism traces tunable trade-offs between welfare, revenue, and fairness as $\lambda$ varies. For RSD, the fairness entry reflects equality of allocation shares; the VCG and learned-mechanism entries are utility-based Jain indices and are therefore not directly comparable.}
\label{tab:airport_results}
\end{table}

\subsection{Empirical Results II: Equity and Efficiency}
\label{subsec:equity_results}

\begin{figure}[t]
    \centering
    \includegraphics[width=\textwidth]{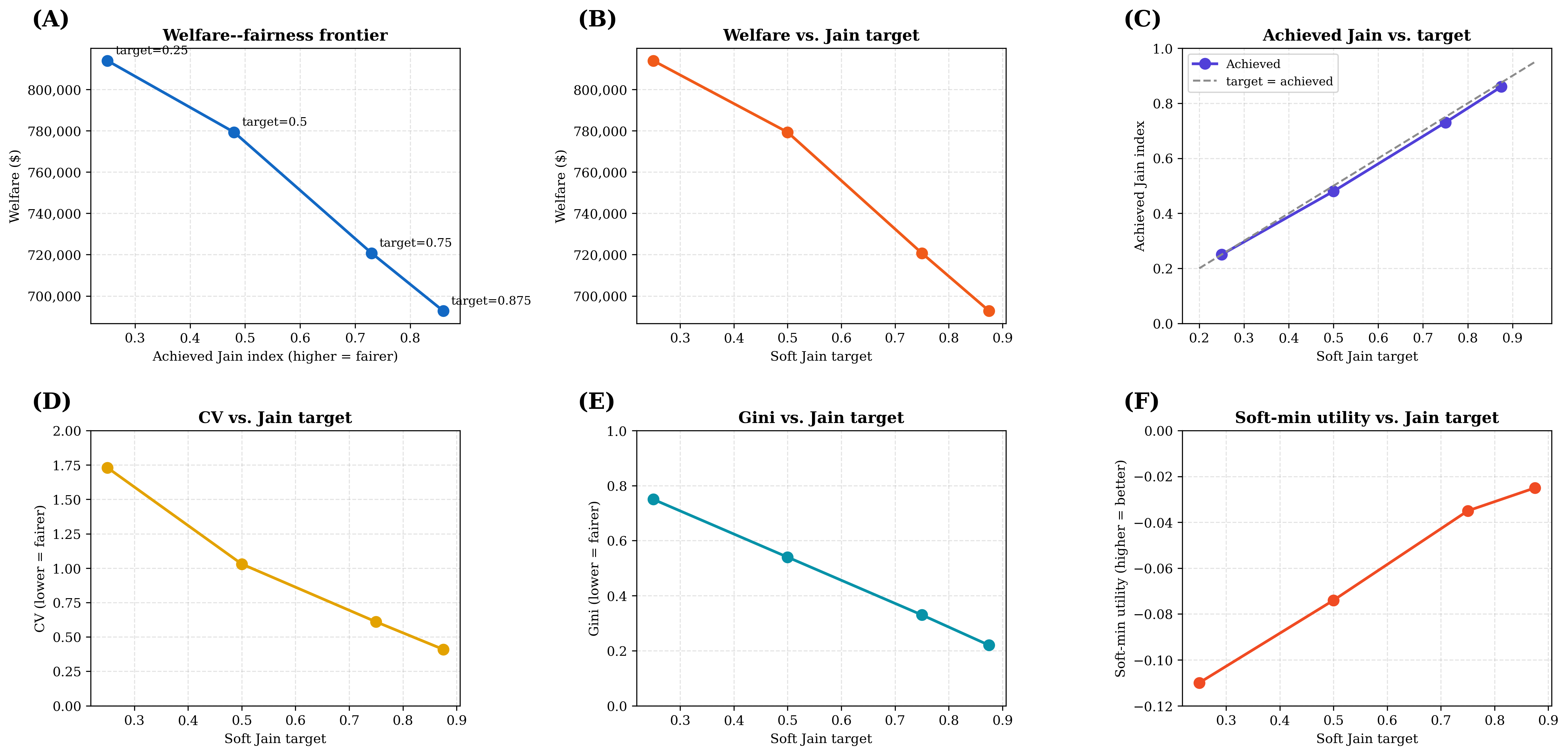}
    \caption{Welfare--fairness trade-offs. \textbf{(A--C)} Pareto frontier and welfare decay relative to target Jain values. \textbf{(D--F)} Auxiliary metrics show how equity changes as the soft targets increase.}
    \label{fig:welfare_fairness}
\end{figure}

In divestiture auctions, regulators must balance allocative efficiency with competitive equity. 

\begin{table}[t]
\centering
\setlength{\tabcolsep}{5pt}
\renewcommand{\arraystretch}{1.1}
\resizebox{\textwidth}{!}{
\begin{tabular}{l|rr|rrrr}
\toprule
\textbf{Metric} & \textbf{RSD} & \textbf{VCG} & \textbf{Target $J{=}0.25$} & \textbf{Target $J{=}0.50$} & \textbf{Target $J{=}0.75$} & \textbf{Target $J{=}0.875$} \\
\midrule
Welfare (\$) & 518,000 & 819,620 & 813,929 & 779,292 & 720,709 & 692,753 \\
Achieved Fairness & 1.00 & 0.27 & 0.25 & 0.48 & 0.73 & 0.86 \\
CV & 0.00 & 1.65 & 1.73 & 1.03 & 0.61 & 0.41 \\
Gini & 0.00 & 0.73 & 0.75 & 0.54 & 0.33 & 0.22 \\
Soft-Min Utility & -- & -0.099 & -0.110 & -0.074 & -0.035 & -0.025 \\
\bottomrule
\end{tabular}
}
\vspace{1mm}
\caption{Welfare--fairness trade-offs under welfare maximization ($\lambda=0$). Higher Jain targets improve equity metrics while reducing welfare. For RSD, the distributional metrics summarize equal allocation shares; for VCG and the learned mechanism, they summarize bidder utilities.}
\label{tab:equity_results}
\end{table}

Table~\ref{tab:equity_results} and Figure~\ref{fig:welfare_fairness} show a favorable welfare--fairness frontier. The lowest-target welfare-maximizing model ($J=0.25$) is highly concentrated (Gini = 0.75), while a target of $J=0.5$ reduces Gini to 0.54 with only a 4\% welfare loss. Higher targets continue to improve dispersion metrics, but welfare losses become more pronounced near the target $J=0.875$. Thus, the learned mechanism provides a tunable alternative to VCG, which maximizes welfare but concentrates utility, and RSD, which equalizes access but ignores valuations.

\subsection{Empirical Results III: Revenue--Fairness Trade-offs}
\label{subsec:revenue_fairness}

Section~\ref{subsec:equity_results} examined the welfare cost of fairness. We now consider the revenue-maximizing regime, where airport operators may seek infrastructure funding while maintaining market equity. Table~\ref{tab:revenue_results} shows that the price of fairness is modest: increasing the Jain target from $J=0.25$ to $J=0.875$ reduces revenue by only 4.3\% (\$396k to \$379k), while reducing the Gini coefficient from 0.75 to 0.17. \textcolor{blue}{Together, the two frontiers show that the price of fairness is objective-dependent: the same increase in the soft Jain target costs 14.9\% of welfare when $\lambda=0$, but only 4.3\% of revenue when $\lambda=1$.}

\begin{table}[h]
\centering
\setlength{\tabcolsep}{6pt}
\renewcommand{\arraystretch}{1.1}
\resizebox{0.8\textwidth}{!}{
\begin{tabular}{lrrrrr}
\toprule
\textbf{Target} & \textbf{Revenue (\$)} & \textbf{Welfare (\$)} & \textbf{Achieved Fair} & \textbf{CV} & \textbf{Gini} \\
\midrule
$J=0.25$ & 396,103 & 423,713 & 0.25 & 1.73 & 0.75 \\
$J=0.50$ & 382,777 & 382,784 & 0.50 & 1.00 & 0.50 \\
$J=0.75$ & 380,960 & 381,474 & 0.75 & 0.58 & 0.27 \\
$J=0.875$ & 379,054 & 379,313 & 0.87 & 0.38 & 0.17 \\
\bottomrule
\end{tabular}
}
\vspace{1mm}
\caption{Revenue--fairness trade-offs under revenue maximization ($\lambda=1$). Higher Jain targets substantially improve dispersion metrics with modest revenue loss.}
\label{tab:revenue_results}
\end{table}

As fairness targets increase, the revenue-maximizing mechanism extracts surplus while equalizing bidder utilities. This cautions against using $\lambda=1$ without additional surplus-preservation constraints. The learned allocation and payment rules also exhibit revenue-screening behavior: allocations and payments shift toward stronger bidders as revenue weight increases. This resembles virtual-value screening, but should be interpreted as an empirical pattern rather than exact Myerson ironing in the combinatorial approximately-IC setting.

\begin{figure}[t]
    \centering
    \includegraphics[width=\textwidth]{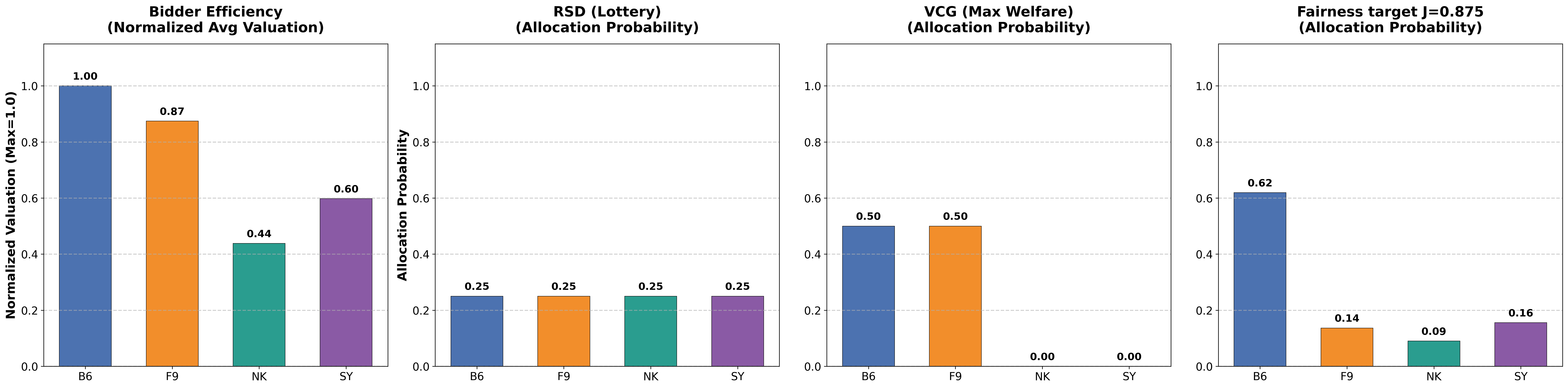}
    \caption{Bidder efficiency and allocation outcomes. \textbf{(a)} Normalized average valuations. \textbf{(b)} RSD provides equal shares regardless of efficiency. \textbf{(c)} VCG produces concentrated allocations. \textbf{(d)} The mechanism trained with target $J=0.875$ redistributes probability toward entrants while preserving efficiency through B6.}
    \label{fig:allocation_comparison}
\end{figure}

Figure~\ref{fig:allocation_comparison} visualizes the resulting redistribution. RSD equalizes shares without using valuation information, whereas VCG concentrates allocations among high-value carriers. Under the higher fairness target, the learned mechanism redistributes probability toward entrants (NK, SY) while preserving efficiency by assigning a larger share to the most productive carrier, B6. Entrant SY receives a slightly higher allocation probability than F9 despite lower average valuation, reflecting the fact that less efficient carriers require more allocation probability to reach comparable utility.

\subsection{Incentive Compatibility Audit}

\begin{table}[h]
\centering
\small
\caption{Aggregate empirical regret audit for airport-slot experiments. Regret is the mean validation regret from adversarial misreport optimization. Detailed per-configuration values are in Appendix~\ref{app:regret_audit}.}
\label{tab:airport_regret_audit_main}
\begin{tabular}{lccc}
\toprule
Frontier & Revenue gain (\$k) & Max regret & Max regret/gain \\
\midrule
Revenue--welfare & 11.2--252.4 & 0.003304 & 0.001750 \\
Welfare--fairness & 11.4--252.6 & 0.009387 & 0.004900 \\
Revenue--fairness & 240.1--257.1 & 0.002331 & 0.000057 \\
\bottomrule
\end{tabular}
\end{table}

We audit approximate incentive compatibility using the aggregate validation regret produced by the adversarial misreport optimization procedure. For each validation profile, candidate misreports are initialized and optimized by gradient ascent according to the saved validation configuration; the largest utility improvement found by that search is reported. As in prior differentiable mechanism design work, this regret is an empirical local-best-response estimate. Table~\ref{tab:airport_regret_audit_main} summarizes the maximum mean regret observed across each reported frontier. In all three families, empirical regret is small relative to the revenue gains obtained over VCG. These values audit the learned fractional mechanism. They do not by themselves establish low regret after discrete resolution and conditional pricing.

\subsection{Empirical Results IV: Ex-Post Outcomes and Practical Implementation}
\label{subsec:practical_implementation}

The frontiers above are ex-ante quantities computed from the learned fractional allocation and payment rules. To implement the mechanism, we resolve the fractional allocation into a feasible deterministic schedule via the lottery procedure in Appendix~\ref{app:decomposition}. Because this introduces sampling variation, realized ex-post outcomes may differ from ex-ante expectations. We quantify this implementation risk using 1{,}000 lottery draws across 20{,}000 holdout valuation instances.

\paragraph{Decomposition Audit.}
As a reproducibility check, we independently retrained a seed-0 CAGraph model using the checked-in airport data and configuration and decomposed all 20{,}000 holdout allocations. The normalization factor was $c_T=1$ for every profile. Mean and maximum $\lVert E_T\rVert_1$ were $3.81\times10^{-15}$ and $1.25\times10^{-14}$, respectively, and the largest $v_{\max}\lVert E_T\rVert_1$ was below $1.50\times10^{-8}$ dollars. The mean support size was 15.5 schedules and the maximum was 28. Appendix~\ref{app:decomposition_audit} reports the complete audit and its provenance. This structural rerun is separate from the economic estimates reported below.

\paragraph{Welfare Stability.}
Allocative efficiency is stable across draws. Under a welfare-maximizing ($\lambda=0$) configuration, the mechanism achieves mean social welfare of \$816{,}527 with standard deviation \$12{,}028. Across draws, 100\% of realized outcomes fall within 20\% of the continuous ex-ante target, and 54.5\% fall within 5\%. This suggests that the learned fractional allocations are close to high-confidence deterministic schedules.

\paragraph{Revenue Reliability.}
Revenue is more volatile because proportional ex-post pricing imposes a no-entrance-fee condition: bidders who receive no bundle pay zero. The learned mechanism has ex-ante expected revenue of \$390{,}357, compared with \$139{,}599 for VCG. In this experiment, the standard lottery approximately preserves the target on average, with realized mean revenue of \$390{,}605, but has a wide lower tail: the 25th percentile is zero because some draws produce no paying winner. The full-assignment variant eliminates this lower-tail risk by resampling until a nonempty feasible assignment is obtained.

\vspace{-8pt}
\begin{table}[h]
\centering
\setlength{\tabcolsep}{7pt}
\renewcommand{\arraystretch}{1.05}
\resizebox{0.82\textwidth}{!}{
\begin{tabular}{lrrrrr}
\toprule
\textbf{Metric} & \textbf{Mean (\$)} & \textbf{Std. Dev (\$)} & \textbf{25th Pct (\$)} & \textbf{Median (\$)} & \textbf{75th Pct (\$)} \\
\midrule
\textbf{Revenue (VCG Baseline)} & 139,599 & 25,608 & 122,366 & 141,425 & 158,656 \\
\textbf{Revenue (Ex-Ante Expected)} & 390,357 & 78,168 & 336,750 & 374,910 & 439,440 \\
\textbf{Revenue (Lottery)} & 390,605 & 283,853 & 0 & 429,735 & 548,226 \\
\textbf{Revenue (Lottery, Full Assignment)} & \textbf{536,328} & 180,213 & \textbf{411,399} & \textbf{487,287} & \textbf{647,464} \\
\bottomrule
\end{tabular}
}
\vspace{1mm}
\caption{Empirical distribution of ex-post revenue over 1{,}000 lottery draws. The standard lottery approximately preserves ex-ante expected revenue in this experiment, while full assignment increases mean revenue by 284.2\% relative to VCG.}
\label{tab:empirical_bounds}
\end{table}
\vspace{-6pt}

\begin{figure}[h]
    \centering
\includegraphics[width=0.9\textwidth]{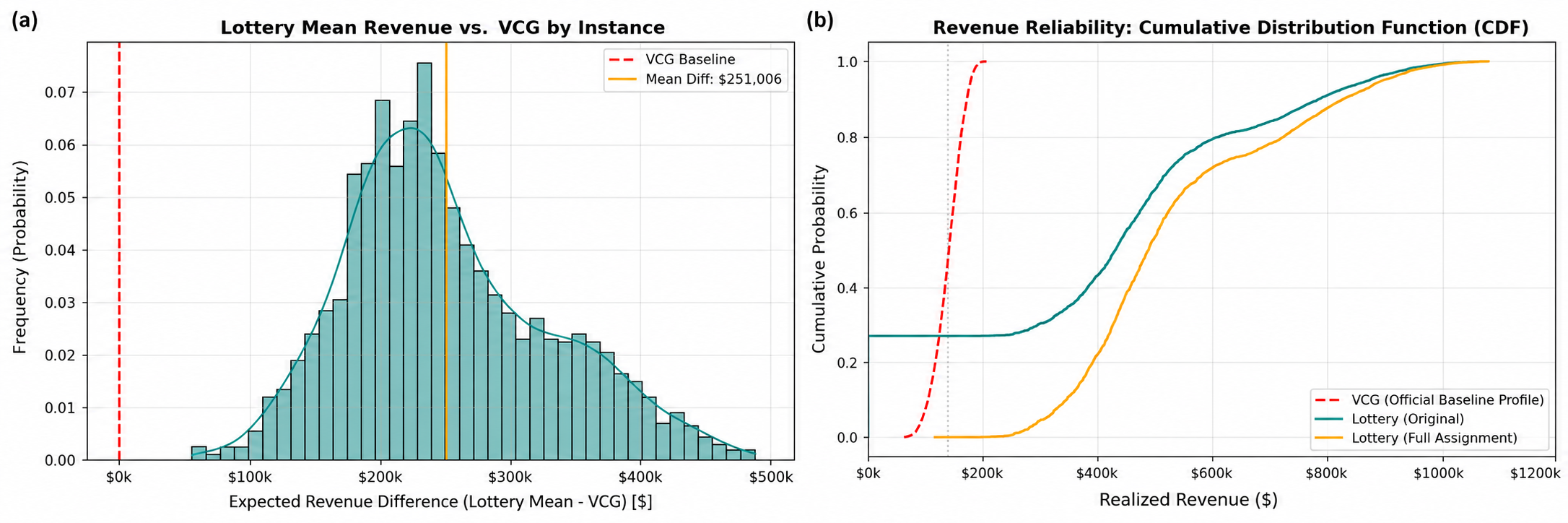}
    \caption{Economic performance. \textbf{(a)} Expected revenue difference distribution. \textbf{(b)} Revenue CDF comparing VCG, standard lottery, and full-assignment implementation.}
    \label{fig:revenue_performance}
\end{figure}

\paragraph{Full-Assignment Implementation.}
As shown in Table~\ref{tab:empirical_bounds} and Figure~\ref{fig:revenue_performance}, full assignment improves both mean revenue and downside protection. Mean revenue rises to \$536{,}328, a 284.2\% increase over VCG, and its 25th percentile (\$411{,}399) exceeds the 75th percentile of VCG (\$158{,}656). Thus, full assignment provides a practical implementation rule that preserves feasibility, avoids entrance fees, and delivers more reliable revenue than the standard lottery. Because conditioning on a nonempty outcome changes the lottery distribution, it also changes expected allocation, revenue, and potentially regret; it should therefore be treated as a separate implementation variant.

\subsection{Structural Robustness and Scalability}
\label{subsec:sensitivity}

\begin{figure}[t]
    \centering
    \includegraphics[width=\textwidth]{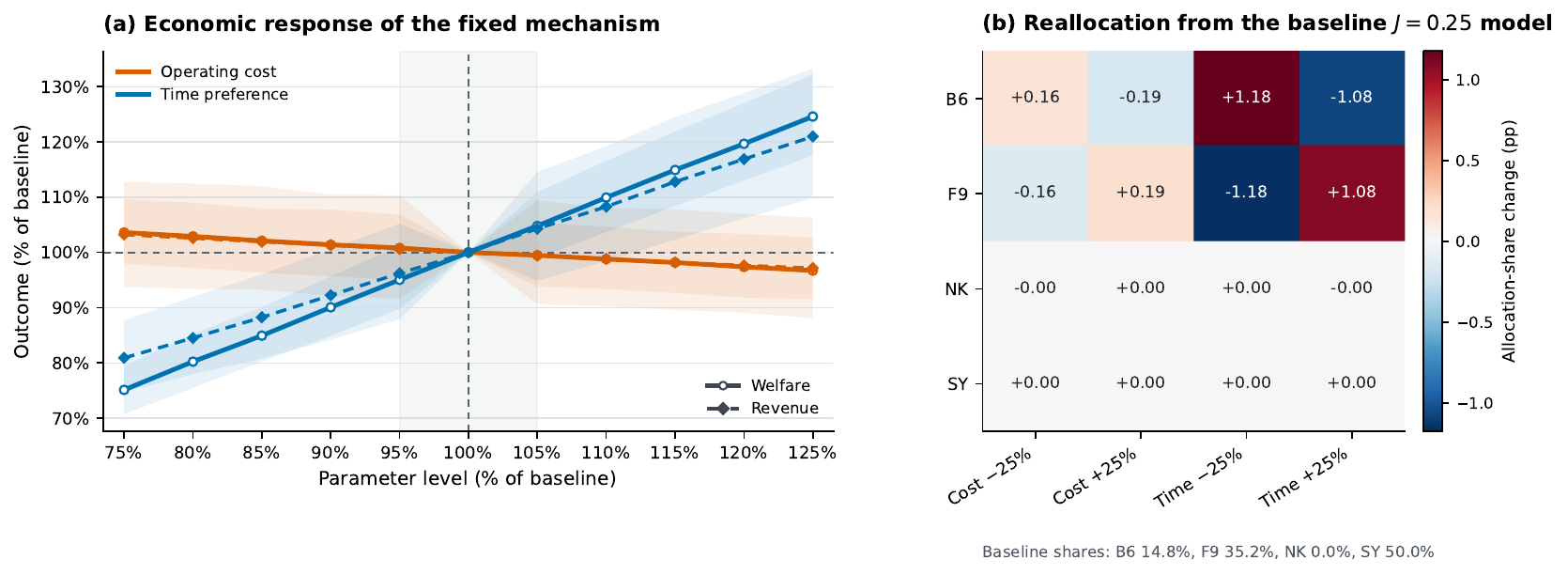}
    \caption{One-at-a-time sensitivity of the fixed trained mechanism. \textbf{(a)} Welfare and revenue are indexed to their parameter-specific baseline means; shaded bands are 95\% nonparametric bootstrap intervals over the 100 regenerated profiles at each level. \textbf{(b)} Endpoint changes in bidder-level allocation shares, in percentage points, relative to the corresponding baseline evaluation. The sensitivity checkpoint uses the welfare--fairness objective with target $J=0.25$ and differs from the target-$J=0.875$ mechanism in Figure~\ref{fig:allocation_comparison}(d).}
    \label{fig:sensitivity_results}
\end{figure}

\textit{Parameter sensitivity.}
Because the airport valuations are model-based estimates, we regenerate bundle valuations after perturbing the calibrated operating-cost coefficient and time-of-day preference levels one at a time from $75\%$ to $125\%$ of baseline. Each level contains 100 independently regenerated profiles, and the fixed trained mechanism is evaluated using its training-data normalization and bidder-level dollar-payment conversion. Figure~\ref{fig:sensitivity_results} shows distinct responses. Operating-cost changes are economically modest: at the $\pm25\%$ endpoints, welfare changes by $+3.6\%/-3.3\%$ and revenue by $+3.2\%/-2.8\%$, with local point elasticities of $-0.13$ and $-0.12$, respectively. Time-of-day preferences are more consequential: the corresponding welfare changes are $-24.9\%/+24.6\%$ and the revenue changes are $-19.1\%/+21.0\%$, with local point elasticities of $0.98$ and $0.81$. The bootstrap bands show that the cost response is small relative to Monte Carlo uncertainty, whereas the endpoint time-preference effects remain separated from baseline. Reallocation remains concentrated between B6 and F9: the $\pm25\%$ time perturbations move their allocation shares by about 1.1--1.2 percentage points in opposite directions, whereas NK and SY remain unchanged at the reported precision. The shaded bands are 95\% nonparametric bootstrap intervals and appropriately reflect that the profiles are independently regenerated at each level. These results use the welfare--fairness checkpoint with target $J=0.25$; this differs from the target-$J=0.875$ mechanism in Figure~\ref{fig:allocation_comparison}(d).


\textit{Allocation stability.}
We measure allocation stability by $\|Z(\epsilon)-Z(0)\|_F$, the Frobenius distance between perturbed and baseline allocation matrices. Even under $\pm 25\%$ perturbations, this distance remains below 0.08, with changes concentrated on marginal bundles rather than core allocations.

\textit{Computational scalability.}
On an NVIDIA Quadro RTX 8000, a forward pass, including winner determination and pricing, takes under 15ms. Complexity analysis is presented in Appendix~\ref{app:complexity}. For larger slot systems, Appendix~\ref{app:scaling} outlines a possible hierarchical decomposition that would partition the conflict graph into weakly coupled temporal submarkets, solve local combinatorial auctions in parallel, and reconcile inter-block dependencies through a coordination layer. This extension is proposed for future work and is not evaluated in the present experiments.

\subsection{Policy Implications and Practical Feasibility}

The transition from traditional, rigid mechanisms to differentiable combinatorial auctions offers three primary strategic advantages for infrastructure managers and regulators:

\paragraph{Auditing Incentive Trade-offs} Reporting empirical regret alongside revenue and welfare allows managers to compare economic gains with the largest deviations found by the configured misreport search. A regulator may use this evidence to screen candidate mechanisms. By accepting a regret budget, regulators can unlock higher revenue or welfare in complex markets, transforming incentive compatibility from a binary constraint into a manageable economic variable for the public interest.

\paragraph{Policy Validation Support} While neural mechanisms are trained for specific market distributions, their sub-15ms inference enables rapid re-evaluation as bids change within the trained market specification. Structural changes to the feasible bundles or conflict graph require a separate validation or retraining. The model may therefore supplement, rather than replace, transparent policies by providing rapid checks of candidate trade-offs.

\paragraph{Flexible Policy Instrument} Our multi-objective framework introduces flexible policy via the weighting factor $\lambda$ or the Jain-index target $J$. By tuning these quantities, managers can prioritize different objectives, transforming a fixed auction into a tunable policy instrument.


\section{Conclusion}
\label{sec:conclusion}

Beyond extending differentiable economics to the combinatorial domain, this paper presents an airport-slot case study and addresses the ``last-mile'' problem of converting fractional combinatorial allocations into feasible discrete schedules. CAGraph models bundle conflicts while tracing the revenue--welfare--fairness frontier, and the Matching Pursuit-inspired decomposition provides a tractable discrete approximation whose residual quantifies the remaining implementation gap.

Our empirical case study on airport slot allocation demonstrates that this integrated approach—combining CAGraph architectures with discrete implementation logic—allows regulators to explicitly navigate the revenue--welfare--fairness Pareto frontier with unprecedented precision. We find expected revenue gains of up to 73.5\% in synthetic settings and 180\% in the airport-slot case study compared to VCG, while maintaining low empirical regret. Ultimately, this framework transforms auction design from a rigid mathematical exercise into a flexible, data-driven policy instrument for modern resource management.

Beyond revenue, the mechanism's millisecond inference enables agile management, such as intra-day reallocation. While our models are currently distribution-specific, future work in zero-shot transfer, or meta-learning could adapt these architectures to structural market changes without retraining. By treating incentive alignment as a manageable economic variable and providing the tools to visualize its impact on policy goals, this work offers a practical, high-performance path for modernizing large-scale resource allocation.

\clearpage

\bibliographystyle{plainnat}
\bibliography{sample}

\clearpage
\appendix
\renewcommand*{\theHsection}{appendix.\arabic{section}}
\renewcommand*{\theHsubsection}{appendix.\arabic{section}.\arabic{subsection}}

\section{Notation Summary}

\begin{table}[H]
    \centering
    \caption{Comprehensive Notation Summary}
    \label{tab:notation}
    \begin{tabular}{ll}
        \hline
        \textbf{Symbol} & \textbf{Definition} \\
        \hline
        $n,m,k$ & Number of bidders, items, and candidate bundles \\
        $N,M,K$ & Sets of bidders, items, and candidate bundles \\
        $v_i, b_i$ & True valuation and reported bid of agent $i$ \\
        $v, b$ & Profiles of valuations and bids for all agents \\
        $V_i, F_i$ & Support and distribution of valuations for agent $i$ \\
        $g(\cdot), p(\cdot)$ & Allocation and payment rules of the mechanism \\
        $u_i(v_i;b)$ & Quasi-linear utility under report profile $b$ \\
        $\operatorname{rgt}_i$ & Empirical ex-post regret for bidder $i$ \\
        $z_{i,S}$ & Probability that agent $i$ receives bundle $S$ \\
        $Z$ & Fractional allocation matrix (output of neural net) \\
        $D_t$ & Feasible deterministic schedule selected at iteration $t$ \\
        $\Omega_t$ & Positive entries selected by schedule $D_t$ \\
        $\gamma_t$ & Nonnegative raw extraction weight \\
        $R_t$ & Raw extraction residual, with $R_0=Z$ \\
        $A_T$ & Raw sparse approximation $\sum_{t<T}\gamma_tD_t$ \\
        $c_T$ & Normalizer $\max\{1,\sum_{t<T}\gamma_t\}$ \\
        $\pi_t$ & Lottery weight $\gamma_t/c_T$ \\
        $\widehat Z_T$ & Expected allocation of the implemented lottery \\
        $E_T$ & Implementation residual $Z-\widehat Z_T$ \\
        $v_{\max}$ & Common upper bound on bundle valuations \\
        $\operatorname{Jain}(u)$ & Jain's fairness index of utilities \\
        $\operatorname{Rev}$ & Total auction revenue $\sum_{i\in N}p_i(v)$ \\
        $w_{\mathrm{rgt}},w_{\mathrm{fair}}$ & Adaptive regret and fairness weights \\
        \hline
    \end{tabular}
\end{table}

\section{Algorithms}

\begin{algorithm}[tb]
    \caption{Adam-Style Update for an Adaptive Penalty Weight}
    \label{alg:update_rule}
    \begin{algorithmic}[1]
        \State Initialize moment states $m_0\gets0$, $s_0\gets0$ and weight $w_0\geq0$
        \For{update $t=1,2,\ldots$}
            \State Compute a signed violation signal $g_t$ relative to the target
            \State $m_t\gets\beta_1m_{t-1}+(1-\beta_1)g_t$
            \State $s_t\gets\beta_2s_{t-1}+(1-\beta_2)g_t^2$
            \State $\widehat m_t\gets m_t/(1-\beta_1^t)$,
                   $\widehat s_t\gets s_t/(1-\beta_2^t)$
            \State $\Delta_t\gets\eta_w\widehat m_t/(\sqrt{\widehat s_t}+\epsilon)$
            \State $w_t\gets\operatorname{clip}(w_{t-1}+\Delta_t,w_{\min},w_{\max})$
        \EndFor
        \State \Return $w_t$
    \end{algorithmic}
\end{algorithm}

\begin{algorithm}[tb]
    \caption{Adaptive Weight Training Algorithm}
    \label{alg:training_appendix}
    \begin{algorithmic}[1]
        \State \textbf{Initialize:} Networks parameters $\theta$, adaptive weights $w_{rgt}, w_{fair}$, targets.
        \For{iteration $t = 1$ to $T$}
            \State Sample batch of valuations $v \sim F$.
            \State Compute misreports $v'$ using gradient ascent on $u(v', v_{-i})$.
            \State Forward pass: Compute allocation $g(v)$, payments $p(v)$.
            \State Compute Objectives: Revenue, Welfare $\mathcal{O}$.
            \State Compute Violations: Regret $G_{rgt}$, Fairness $G_{fair}$.
            \State Compute Loss: $\mathcal{L} = -\log(1 + \mathcal{O}) + w_{rgt} G_{rgt} + w_{fair} G_{fair}$.
            \State Update network parameters $\theta \leftarrow \theta - \eta \nabla_\theta \mathcal{L}$.
            \State \textbf{Update Adaptive Weights:}
            \State \quad $g_{rgt} \leftarrow \log(G_{rgt}) - \log(\text{target}_{rgt})$
            \State \quad $w_{rgt} \leftarrow \text{AdamUpdate}(w_{rgt}, g_{rgt})$
            \State \quad Decentralize target: $\text{target}_{rgt} \leftarrow \text{decay}(\text{target}_{rgt})$.
        \EndFor
    \end{algorithmic}
\end{algorithm}





\section{Training Procedure and Loss Function}\label{app:training}

This appendix describes the full training procedure for learning a fair,
approximately-truthful deep auction mechanism. The mechanism consists of an
allocation network and a payment network, denoted jointly by $M_\theta$, which
map reported valuations $x\in\mathbb{R}^{A\times M}$ and context $c$ to
allocations $\alpha_\theta(x,c)$ and payments $p_\theta(x,c)$.

Training balances three goals: (i) efficiency, (ii) incentive alignment via low
regret, and (iii) fairness of realized utilities.

\subsection{Warm-Start via VCG Imitation}

For initialization, the mechanism is trained to imitate the VCG outcome.
Given $(x,c)$ we compute
\begin{equation}
(\alpha^{\mathrm{VCG}},\, p^{\mathrm{VCG}})
\;\leftarrow\;
\mathrm{VCG}(x,c).
\end{equation}

The warm-start loss is
\begin{align}
L_{\mathrm{warm}}(\theta)
  &= \,L_{\alpha}(\theta)
   + \,L_{p}(\theta), \text{ where}
\\[3pt]
L_{\alpha}(\theta)
  &= \|\alpha_\theta(x,c)
      - \alpha^{\mathrm{VCG}}\|_2^{2}, \text{ and}
\\[3pt]
L_{p}(\theta)
  &= \|p_\theta(x,c)
      - p^{\mathrm{VCG}}\|_2^{2}.
\end{align}

\subsection{Adversarial Misreport Search}

Misreports are parameterized as
\begin{equation}
\tilde{x} = f_{\mathrm{adv}}(x;\phi),
\end{equation}
where $\phi$ is an adversarial tensor optimized by gradient descent.

Truthful and misreport utilities use true valuations:
\begin{equation}
u_i(x)
  = \bigl\langle \alpha_{\theta,i}(x,c),\, x_i \bigr\rangle
    - p_{\theta,i}(x,c).
\end{equation}

The adversarial objective (one step of inner maximization) is
\begin{equation}
L_{\mathrm{adv}}(\phi)
   = -\sum_{i=1}^{A} u_i(\tilde{x}_i;\, x).
\end{equation}

\subsection{Regret Computation}

Ex-post regret is
\begin{equation}
\mathrm{Regret}(x;\theta)
  = \frac{1}{A}
    \sum_{i=1}^{A}
     \Bigl[
       \max_{\tilde{x}_i}
         \bigl( u_i(\tilde{x}_i;x)
              - u_i(x_i;x) \bigr)
     \Bigr]_{+}.
\end{equation}

The inner maximization is approximated numerically by optimizing initialized
candidate reports while holding the other bidders' reports fixed. Consequently,
the reported quantity is empirical regret: additional starts or optimization
steps can reveal a larger profitable deviation.

\subsection{Payment Parameterization and Truthful IR}

The payment head outputs a fraction $q_i(b)\in[0,1]$ and sets
\begin{equation}
p_i(b)=q_i(b)\sum_{S\in K}g_{iS}(b)b_{iS}.
\end{equation}
At a truthful report $b_i=v_i$, the fractional mechanism therefore satisfies
\begin{equation}
u_i(v_i;v)=
\bigl(1-q_i(v)\bigr)\sum_{S\in K}g_{iS}(v)v_{iS}\geq0.
\end{equation}
This is pointwise truthful-profile IR for the fractional expected outcome
(that is, before the lottery is realized). The separate
conditional payment rule used after discrete resolution is described in
Section~\ref{app:decomposition}; incentive and IR claims for that transformed
mechanism must be assessed at the post-resolution level.

\subsection{Fairness Metrics}

Truthful expected utilities are
\begin{equation}
u_i = \mathbb{E}\!\left[
      (\alpha_{\theta,i}(x,c)\cdot x_i)
      - p_{\theta,i}(x,c)
    \right].
\end{equation}

We monitor several fairness indices for evaluation:
\begin{align}
\mathrm{Jain}(u)
 &= \frac{(\sum_i u_i)^2}
         {A \sum_i u_i^2},
\\[2mm]
\mathrm{Gini}(u)
 &= \frac{
     \sum_{i}\sum_{j} |u_i - u_j|
   }{
     2A\sum_i u_i
   },
\\[2mm]
\mathrm{CV}(u)
 &= \frac{\sigma(u)}{\bar{u}},
\\[2mm]
\mathrm{SoftMin}(u)
 &= -\log\!\left(\sum_i e^{-u_i}\right).
\end{align}

For a selected fairness target $J_0$, training uses the soft target penalty
\begin{equation}
L_{\mathrm{fair}}(\theta)
  = \bigl(\mathrm{Jain}(u)-J_0\bigr)^2.
\end{equation}
This is an equality-style regularizer, not a hard lower-bound constraint. The
other indices are evaluation diagnostics.

\subsection{Economic Objective}

Let $\mathrm{Obj}(\theta)$ denote a mixture of revenue and welfare:
\begin{equation}
\mathrm{Obj}(\theta)
  =
\begin{cases}
\mathrm{Revenue}, & \text{``revenue''}, \\[3pt]
\mathrm{Welfare}, & \text{``welfare''}, \\[3pt]
0.5\,\mathrm{Rev}+0.5\,\mathrm{Wel},
   & \text{``50-50''}, \\[3pt]
\text{other mixtures}. &
\end{cases}
\end{equation}

For stability, we optimize the log-transformed loss
\begin{equation}
L_{\mathrm{econ}}(\theta)
   = -\log\!\left(1+\mathrm{Obj}(\theta)\right).
\end{equation}
This only rescales the objective and the gradients, not affecting the non-convexity property.

\subsection{Optional Revenue-Range Regularization}

When a training configuration includes a desired revenue range
$[R_{\min},R_{\max}]$, it applies the soft penalty
\begin{align}
\delta_{\min}&=(R_{\min}-R)_{+},
&\delta_{\max}&=(R-R_{\max})_{+},\\
L_{\mathrm{revbd}}
&=\log\!\left(1+\gamma_{R}
  (\delta_{\min}^{2}+\delta_{\max}^{2})\right).
\label{eq:ratio-loss}
\end{align}
The bounds are configuration parameters and may be expressed in the same
normalized or monetary units as the training objective. This term is a
regularizer, not a hard feasibility constraint, and no fixed
revenue-to-welfare percentages are assumed here.

\subsection{Adaptive Weights for Regret and Fairness}

We maintain two adaptive multipliers:
\[
w_{\mathrm{rgt}} \quad\text{and}\quad w_{\mathrm{fair}},
\]
updated via an Adam-style rule.

Define the regret signal in log-space:
\[
g^{(\mathrm{rgt})}_t
= 
\log\!\left( 
\frac{\mathrm{Regret}}{\text{rgt\_ratio}\cdot \mathrm{Obj} + 1}
\right)
-
\log(\text{rgt\_target}).
\]

For fairness, let $\varepsilon_{\mathrm{fair}}=1-J_0$. The signed update
signal is
\[
g^{(\mathrm{fair})}_t
=
\bigl(1-\mathrm{Jain}(u)\bigr)-\varepsilon_{\mathrm{fair}},
\]
while the mechanism loss uses the square of this gap.

Each weight is updated via:
\[
\begin{aligned}
m_t &= 0.9 m_{t-1} + 0.1 g_t, \\
v_t &= 0.999 v_{t-1} + 0.001 g_t^2, \\
w_t &= \mathrm{clip}\!\left(
w_{t-1} + \eta \frac{m_t}{\sqrt{v_t}+10^{-8}},\,
0.1,\,
5.0
\right).
\end{aligned}
\]

\subsection{Full Training Loss}

The complete loss for updating the mechanism parameters $\theta$ is:
\[
\boxed{
L(\theta)
=
L_{\mathrm{econ}}
+
w_{\mathrm{rgt}} \cdot \mathrm{Regret}
+
w_{\mathrm{fair}} \cdot L_{\mathrm{fair}}
+
L_{\mathrm{revbd}}.
}
\]

\subsection{Interpretation of High-Revenue Solutions}

In some runs, the learned payment rule produces revenue close to allocated
welfare. This is an empirical training outcome, not a consequence of a general
VCG payment-minimality theorem. The allocation and payment networks are jointly
optimized under finite-sample regret penalties and soft regularization, so a
high-revenue solution can reflect the selected economic objective, the payment
parameterization, approximation error in the inner misreport search, and the
strength of the penalty weights. We therefore interpret revenue, regret, and IR
together on held-out data rather than labeling such a solution DSIC from its
revenue level alone.

\paragraph{Notation.} Let $\theta$ denote the parameters of the mechanism $M_\theta$. Given reports $x \in \mathbb{R}^{A \times M}$ and context $c$, the mechanism outputs allocations $\alpha_\theta(x,c)$ and payments $p_\theta(x,c)$. For agent $i$, $u_i(x,c; \theta) = \alpha_{\theta,i}(x,c) \cdot x_i - p_{\theta,i}(x,c)$ is the utility. Welfare and revenue are defined as:
\begin{align}
    W(\theta) &= \mathbb{E}_{(x,c) \sim \mathcal{D}} \left[ \sum_{i=1}^A \alpha_{\theta,i}(x,c) \cdot x_i \right] \\
    R(\theta) &= \mathbb{E}_{(x,c) \sim \mathcal{D}} \left[ \sum_{i=1}^A p_{\theta,i}(x,c) \right]
\end{align}

The regret at $(x,c)$ is:
\begin{align}
    \text{Regret}(x,c; \theta) = \frac{1}{A} \sum_{i=1}^A \max_{\tilde{x}_i} \bigl[ &u_i(\tilde{x}_i,x_{-i},c; \theta) \nonumber \\ 
    &- u_i(x_i,x_{-i},c; \theta) \bigr]_+
\end{align}

and its expectation is $\text{Reg}(\theta) = \mathbb{E}_{(x,c) \sim \mathcal{D}} [\text{Regret}(x,c; \theta)]$.

Let $\bar{u}_i(\theta) = \mathbb{E}_{(x,c) \sim \mathcal{D}} [u_i(x,c; \theta)]$ and $u(\theta) = (\bar{u}_1(\theta), \dots, \bar{u}_A(\theta))$. The Jain index of $u(\theta)$ is:
\begin{equation}
    \text{Jain}(u(\theta)) = \frac{\left( \sum_{i=1}^A \bar{u}_i(\theta) \right)^2}{A \sum_{i=1}^A \bar{u}_i(\theta)^2}
\end{equation}

\paragraph{Economic objective.} For a specified mixture parameter $\lambda \in [0,1]$, the objective is:
\begin{equation}
    \text{Obj}_\lambda(\theta) = \lambda \text{Rev}(\theta) + (1-\lambda) \text{Wel}(\theta)
\end{equation}

\paragraph{Soft-penalty implementation.} The training code does not solve a
hard constrained program. Instead, it maximizes
$\operatorname{Obj}_\lambda$ through the log-transformed economic loss and
uses adaptive penalties to target low regret and the selected Jain index. When
enabled, the revenue-range term in \eqref{eq:ratio-loss} is also soft. Thus,
reported target levels should be interpreted as optimization targets whose
attainment is checked empirically, not as feasibility guarantees.

Primary airport economic outcomes are reported in the main empirical tables so
that a single set of point estimates is authoritative. The appendix reports
methodological detail and the regret audit derived from those same reported
configurations, rather than an earlier validation snapshot.

This section provides the formal mathematical derivations for the structural guarantees and implementation theorems presented in the main text.

\section{Proof of Theorem~\ref{thm:feasibility} and Implementation Details}
\label{app:feasibility_proof}

Let $K$ be the set of feasible bundles and let $k:=|K|$.
Our layer outputs two nonnegative matrices:
\[
p\in[0,1]^k\quad\text{(``bundle availability'')}
\]
\[
q\in[0,1]^{n\times k}\quad\text{(``bundle-to-agent assignment'')}.
\]
They are produced as follows.
For each item $j\in M$, the masked normalization over bundle logits yields
nonnegative per-item probabilities $\tilde p_{j,S}$ satisfying
$\sum_{S\in K:S\ni j}\tilde p_{j,S}\leq1$. Equality holds when the
normalization is restricted exactly to bundles containing $j$.
We set
\[
p_S \;=\; \min_{j\in S}\,\tilde p_{j,S}
\]
\[\Rightarrow\quad
\forall j\in M:\ \sum_{S\in K:S\ni j} p_S
\le \sum_{S\in K:S\ni j}\tilde p_{j,S}\leq1.
\]
For the assignment, we compute $q$ as the elementwise minimum of two softmax normalizations:
\[
q \;=\; \min\!\bigl(\text{softmax}_{\text{agents}}(A), \text{softmax}_{\text{bundles}}(A')\,\bigr).
\]
Hence $q$ is \emph{doubly sub-stochastic}:
\[
\forall S\in K:\ \sum_{i=1}^n q_{i,S}\le 1,
\qquad
\forall i\in N:\ \sum_{S\in K} q_{i,S}\le 1.
\]
The final allocation is
\[
z_{i,S} \;=\; p_S\,q_{i,S}\in[0,1].
\]
It satisfies bidder feasibility because
\[
\sum_{S\in K}z_{i,S}
\leq\sum_{S\in K}q_{i,S}\leq1.
\]
For any item $j$, bundle feasibility gives
\[
\sum_{i\in N}\sum_{S\in K:S\ni j}z_{i,S}
=\sum_{S\in K:S\ni j}p_S\sum_{i\in N}q_{i,S}
\leq\sum_{S\in K:S\ni j}p_S\leq1.
\]
Thus the layer satisfies both fractional feasibility constraints claimed in
Theorem~\ref{thm:feasibility}.

\section{Revenue--Welfare Weighting}
\label{app:rev_welfare_weighting}

For a fixed mixture parameter $\lambda\in[0,1]$, define the batch economic
objective
\begin{equation}
\label{eq:app_rw_max}
\operatorname{Obj}_{\lambda}(\theta)
=\lambda\operatorname{Rev}_{\theta}
+(1-\lambda)\operatorname{Wel}_{\theta}.
\end{equation}
The implemented economic loss is
\begin{equation}
\label{eq:app_rw_obj}
L_{\mathrm{econ}}
=-\log\!\left(1+\operatorname{Obj}_{\lambda}(\theta)\right).
\end{equation}
Ignoring the penalty terms, the logarithm is monotone and therefore has the
same batchwise optimizer as maximizing $\operatorname{Obj}_{\lambda}$, while
rescaling its gradient. Thus, $\lambda$ controls the empirical trade-off between
revenue extraction and allocative efficiency. In single-item regular environments, revenue maximization admits a
virtual-value characterization. We use this classical result only as an analogy for the
allocation distortions observed in the learned mechanisms. We do not claim that the trained
combinatorial mechanism admits a closed-form virtual-value representation or performs
exact Myerson ironing.

\section{Implementability and Decomposition Proofs}
\label{app:implementability}

\subsection{Implementability Limitation}

Let $\mathcal{B}$ denote the set of deterministic feasible schedules. Each
$B\in\mathcal{B}$ is a binary matrix satisfying the item-wise and bidder-wise feasibility
constraints. A fractional allocation $Z$ is exactly implementable as a lottery over feasible
deterministic schedules if and only if
\[
Z \in \operatorname{conv}(\mathcal{B}).
\]
Equivalently, there exist weights $\{\pi_B\}_{B\in\mathcal{B}}$ such that
\begin{align}
Z &= \sum_{B\in\mathcal{B}}\pi_B B, \label{eq:lottery_rep}\\
\sum_{B\in\mathcal{B}}\pi_B &= 1, \qquad
\pi_B\ge 0. \label{eq:lottery_weights}
\end{align}

For simple assignment problems, the Birkhoff--von Neumann theorem guarantees this
property because the relevant fractional polytope is integral. However, combinatorial
bundle constraints generally produce set-packing polytopes whose linear relaxations are
not integral. The bihierarchy characterization of \citet{budish2013designing} gives a
structural condition under which universal implementation is possible for canonical
two-sided constraint systems. When the constraint structure is not a bihierarchy, universal
implementation fails: there exist feasible fractional allocations that cannot be expressed as
lotteries over deterministic feasible allocations.

Airport slot bundles can induce precisely this failure. A bundle may contain multiple
physical or temporal resources, and two bundles conflict whenever they share at least one
resource. Such overlaps can create cycles in the conflict structure, violating the nested or
disjoint structure required by a bihierarchy. Therefore, feasibility of $Z$ under the
fractional constraints does not by itself imply $Z\in\operatorname{conv}(\mathcal{B})$.
This proves Proposition~\ref{prop:implementability_limitation}.

\paragraph{Minimal conflict-cycle example.}
Consider three resources $a,b,c$ and three bundles
\[
S_1=\{a,b\},\quad
S_2=\{b,c\},\quad
S_3=\{a,c\}.
\]
Each pair of bundles overlaps, so at most one bundle can be selected in any deterministic
schedule. The fractional point
\[
z_{S_1}=z_{S_2}=z_{S_3}=1/2
\]
satisfies the per-resource constraints because each resource has total load one. However,
the total bundle mass is $3/2$, whereas any deterministic feasible schedule selects at most
one bundle. Hence this fractional point cannot be represented as a convex combination of
deterministic feasible schedules. This illustrates why fractional feasibility is weaker than
lottery implementability in overlapping-bundle environments.

\subsection{Proof of Proposition~\ref{prop:residual_reduction}}\label{app:residual_reduction_proof}

By construction, $\gamma_t=\min_{(i,S)\in\Omega_t}R_{t,iS}$. For entries with
$D_{t,iS}=0$,
\[
R_{t+1,iS}=R_{t,iS}\ge 0.
\]
For entries with $D_{t,iS}=1$, the definition of $\gamma_t$ implies
\[
0\le \gamma_t\le R_{t,iS},
\]
and therefore
\[
R_{t+1,iS}=R_{t,iS}-\gamma_t\ge 0.
\]
Thus $R_{t+1}\ge 0$. Moreover,
\begin{align}
\|R_{t+1}\|_1
&= \sum_{i,S} R_{t+1,iS} \notag\\
&= \sum_{i,S} R_{t,iS}
-\gamma_t \sum_{i,S} D_{t,iS} \notag\\
&= \|R_t\|_1-\gamma_t\|D_t\|_1.
\end{align}
Since $\gamma_t\ge 0$ and $\|D_t\|_1\ge 0$, it follows that
$\|R_{t+1}\|_1\le \|R_t\|_1$. The inequality is strict whenever $\gamma_t>0$ and
$\Omega_t\ne\emptyset$.

\subsection{Economic Discretization Gap}

The extraction procedure first produces the nonnegative sparse approximation
\[
A_T=\sum_{t=0}^{T-1}\gamma_tD_t,
\qquad
R_T=Z-A_T.
\]
The raw extraction weights need not sum to at most one. Accordingly, define
\[
c_T=\max\left\{1,\sum_{t=0}^{T-1}\gamma_t\right\},
\qquad
\pi_t=\frac{\gamma_t}{c_T},
\]
and assign probability
$\pi_{\emptyset}=1-\sum_{t=0}^{T-1}\pi_t$ to the null schedule. These
probabilities are nonnegative and sum to one. The expected implemented allocation
and its implementation residual are therefore
\[
\widehat Z_T=\sum_{t=0}^{T-1}\pi_tD_t=\frac{A_T}{c_T},
\qquad
E_T=Z-\widehat Z_T.
\]
For any valuation profile $v$ satisfying $0\le v_{iS}\le v_{\max}$,
\begin{align}
\big|\langle v,Z\rangle
-\langle v,\widehat Z_T\rangle\big|
&= |\langle v,E_T\rangle| \notag\\
&\le \|v\|_\infty\|E_T\|_1 \notag\\
&\le v_{\max}\|E_T\|_1.
\end{align}
Therefore,
\[
\big|\langle v,Z\rangle
-\langle v,\widehat Z_T\rangle\big|
\le v_{\max}\|Z-\widehat Z_T\|_1.
\]
Notice that $R_T$ certifies how much nonnegative mass remains after the raw
greedy extraction, whereas $E_T$ measures the error of the normalized lottery
that is actually implementable. They coincide only when $c_T=1$.

\section{Case Study Details}\label{appendix:detailedsimulation}

\subsection{Airport Valuation Construction with Gurobi}
\label{app:airport_valuation_gurobi}

The airport valuations are model-based estimates of the maximum operating profit that a carrier can obtain from a candidate slot bundle. The construction uses the DB1B Ticket and Coupon tables. Coupon records are ordered by \texttt{ITIN\_ID}; the first origin and final destination are joined to the
corresponding ticket record to recover each itinerary's endpoints. For the reported experiment, itineraries are retained when either endpoint is Ronald Reagan Washington National Airport (DCA; BTS airport ID 11278). Each retained flight is marked as a departure if DCA is its origin and as an arrival if DCA is its destination. The resulting route, fare, distance, capacity, and
direction fields are used to construct the flight--slot profit coefficients $V_i(f,t)$ described in Section~\ref{subsec:data_formulation}.

The candidate times are 08:00, 10:00, 17:00, and 22:00. Candidate bundles
contain an even number of these times, yielding the six two-slot combinations
and the single four-slot combination, for seven bundles in total. For a flight
$f$, write $F_f$, $S_f$, and $D_f$ for its \texttt{ITIN\_FARE},
\texttt{SEAT\_CAP}, and \texttt{DISTANCE}. For each generated valuation
profile, the reported training artifact fixes the per-mile operating-cost
coefficient at $c_{\mathrm{mile}}=0.019$ and computes
\begin{align}
\operatorname{Revenue}_f
&=0.7F_fS_f,\\
\operatorname{Cost}_f
&=c_{\mathrm{mile}}(D_f+722)(S_f+104),\\
\operatorname{Profit}_f
&=\operatorname{Revenue}_f-\operatorname{Cost}_f.
\end{align}
The load factor and $c_{\mathrm{mile}}$ are fixed in the reported training
artifact. Time-of-day multipliers start from 0.50, 1.00, 0.85,
0.95, 0.85, and 0.75 for the six bins
00:00--06:00, 06:00--09:00, 09:00--15:00, 15:00--18:00,
18:00--21:00, and 21:00--00:00, respectively. For each generated profile,
each bin value is perturbed independently by a uniform draw on
$[-0.5,0.5]$ and truncated below at zero. The coefficient used by Gurobi is
\[
V_i(f,t)=\operatorname{Profit}_f\,\tau_t,
\]
where $\tau_t$ is the sampled multiplier for the bin containing slot $t$.

For every generated profile, carrier $i$, and candidate bundle $S$, we solve a binary assignment model in Gurobi. Let $F_i$ be the carrier's candidate flights and let $T(S)$ be the slots contained in bundle $S$. The implementation creates binary variables $z_{ft}$, equal to one when flight $f$ is assigned to slot $t$, and $w_f$, equal to one when flight $f$ is not operated. It then solves
\begin{align}
v_i(S)=\max_{z,w}\quad
&\sum_{f\in F_i}\sum_{t\in T(S)}V_i(f,t)z_{ft} \label{eq:app_val_objective}\\
\text{s.t.}\quad
&\sum_{f\in F_i}z_{ft}\leq1,
&&\forall t\in T(S), \label{eq:app_val_slot}\\
&\sum_{t\in T(S)}z_{ft}+w_f=1,
&&\forall f\in F_i, \label{eq:app_val_flight}\\
&\sum_{f\in F_i}\sum_{t\in T(S)}
\operatorname{dir}(f)z_{ft}=0, \label{eq:app_val_balance}\\
&z_{ft},w_f\in\{0,1\}. \label{eq:app_val_binary}
\end{align}
Here $\operatorname{dir}(f)=+1$ for a departure and $-1$ for an arrival. Constraint~\eqref{eq:app_val_slot} prevents two flights from occupying the same slot, constraint~\eqref{eq:app_val_flight} assigns each candidate flight to at most one slot or cancels it, and constraint~\eqref{eq:app_val_balance} enforces arrival--departure flow balance. In the implementation, the model is created with \texttt{gurobipy.Model}, both variable families use
\texttt{GRB.BINARY}, and the objective uses \texttt{GRB.MAXIMIZE}. After \texttt{optimize()}, the objective value \texttt{ObjVal} is recorded as $v_i(S)$ only when the solver returns \texttt{GRB.OPTIMAL}.

Solving this model separately for every carrier--bundle pair produces one
four-carrier, seven-bundle valuation profile. Repeating the construction under
the sampled time-preference multipliers produces
the fixed 100,000-profile
valuation artifact used in the experiments; profiles 1--80,000 form the
training set and profiles 80,001--100,000 form the held-out test set. The same
configuration applies after filtering the source itineraries to DCA; the LGA
variant was a later experiment and is not the source of the reported DCA data.

The sensitivity analysis in Section~\ref{subsec:sensitivity} uses the checked-in
parameter-regenerated valuation files. Eleven levels from 0.75 to 1.25 are
evaluated for each parameter, with 100 independently regenerated profiles per
level. In the cost experiment, the coefficient is set to
$c_{\mathrm{mile}}=0.019s$. In the time-preference experiment, the six
calibrated bin means are multiplied by $s$ before applying the generator's
independent $\pm5\%$ relative jitter. Route, fare, capacity, load-factor, and
bundle-optimization inputs are otherwise unchanged.

The fixed target-$J=0.25$ checkpoint is evaluated at every level. All scenarios
use the minimum and maximum from the training valuation artifact rather than
being normalized separately. The payment head's normalized output is converted
to dollars bidder by bidder: its implied fraction of allocated normalized value
is applied to that bidder's allocated dollar value and then summed. Allocation
columns are matched to the corresponding slot sets before welfare and dollar
payments are calculated because the valuation files and the product-based
incident matrix enumerate the six pair bundles in opposite orders. Outcomes are
indexed to the parameter-specific $s=1$ mean. The shaded bands in
Figure~\ref{fig:sensitivity_results} are percentile intervals from 2,000
nonparametric bootstrap resamples; numerator and baseline profiles are resampled
independently because each parameter level contains an independently regenerated
Monte Carlo sample. Local elasticities use the centered $s=0.95$ and $s=1.05$
evaluations.

\begin{table}[h]
\centering
\small
\caption{Recomputed one-at-a-time sensitivity results. Outcomes are indexed to
the corresponding parameter-specific baseline mean. Local elasticities use the
centered 0.95--1.05 evaluations; brackets give 95\% bootstrap intervals.}
\label{tab:sensitivity_audit}
\begin{adjustbox}{max width=\textwidth}
\begin{tabular}{llrrrl}
\toprule
Parameter & Outcome & 75\% level & Baseline & 125\% level & Local elasticity [95\% interval] \\
\midrule
Operating cost & Welfare & 103.60 & 100.00 & 96.70 & $-0.132\;[-0.746,\ 0.466]$ \\
Operating cost & Revenue & 103.23 & 100.00 & 97.16 & $-0.124\;[-1.038,\ 0.827]$ \\
Time preference & Welfare & 75.11 & 100.00 & 124.60 & $0.975\;[0.400,\ 1.569]$ \\
Time preference & Revenue & 80.90 & 100.00 & 120.97 & $0.806\;[-0.129,\ 1.720]$ \\
\bottomrule
\end{tabular}
\end{adjustbox}
\end{table}

\subsubsection{Slot Allocation at U.S. Coordinated Airports: Practice and Precedent}

In the United States, a small number of high-demand airports—including John F. Kennedy International (JFK), LaGuardia (LGA), and Ronald Reagan Washington National (DCA)—are designated as Level 3 coordinated airports under IATA's Worldwide Slot Guidelines (WSG). Slot allocation at these airports is managed by the Federal Aviation Administration (FAA) through a hybrid system based on regulatory oversight and historical usage rights.

\paragraph{Historical Priority and the 80\% Rule}

The cornerstone of current U.S. slot allocation is the “grandfather rights” system, whereby carriers retain priority to operate a slot in future seasons if they use it at least 80\% of the time during the current season. This “use-it-or-lose-it” rule is intended to deter slot hoarding and promote efficient utilization of limited runway capacity. Slots that are underutilized below this threshold are returned to a common pool for reallocation.

\paragraph{Reallocation Mechanisms}

Slots may become available through several channels: voluntary surrender, underutilization (per the 80\% rule), or involuntary divestiture—particularly in the context of antitrust enforcement or airline mergers. For example, during the 2011 Delta–US Airways slot exchange, the DOT required the divestiture of 24 slot pairs at LGA and DCA to preserve market competitiveness. These divested slots were bundled and auctioned to eligible new entrants or limited incumbents, marking a rare instance of competitive slot auction in the U.S.

When slots return to the pool, the FAA typically uses a randomized lottery to determine the order of selection by eligible carriers. This mechanism resembles a Random Serial Dictatorship (RSD): carriers sequentially select individual slots from the available pool without submitting bids or revealing valuations. These procedures are formalized in \href{https://www.law.cornell.edu/cfr/text/14/93.225}{14 CFR §93.225}, which governs slot lotteries at constrained airports.

Historically, slot lotteries have been employed under the FAA’s High Density Rule (HDR), particularly during the 1980s and early 2000s. For example, the FAA issued a Federal Register notice in 2003 clarifying the lottery process at DCA, including eligibility conditions for new entrants and limited incumbents \href{https://www.federalregister.gov/documents/2003/07/18/03-18401/high-density-airports-notice-of-reagan-national-airport-lottery-allocation-procedure}{(link)}. Additionally, the 2015 Slot Management and Transparency Rule reiterated the FAA’s use of randomized procedures to allocate new or temporarily available slots, often with priority given to new entrants \href{https://www.federalregister.gov/documents/2015/01/08/2014-30378/slot-management-and-transparency-for-laguardia-airport-john-f-kennedy-international-airport-and}{(link)}.

While auctions remain rare in U.S. slot allocation, these precedents illustrate that both randomized and market-based mechanisms are feasible within the current regulatory framework. Our case study builds on these insights by evaluating combinatorial allocation approaches under realistic, small-scale scenarios, such as divestiture or reallocation events involving limited bidders and bundles.

\subsubsection{Comparison to Lottery and Serial Dictatorship Mechanisms}

While our mechanism builds on a learned, data-driven allocation and payment rule, it is instructive to compare it with simpler mechanisms commonly used or considered in practice, such as pure lotteries and serial dictatorship. These mechanisms have intuitive appeal but lack important economic guarantees that our approach approximates.

\paragraph{Pure Lottery.} In the stylized pure-lottery benchmark used in the
main text, bidders receive equal allocation shares without regard to their
valuations. This provides procedural symmetry but can be inefficient and yields
no revenue. Because the benchmark solicits no reports, report-based incentive
compatibility is vacuous; its equal shares should not be interpreted as equal
utility.

\paragraph{Serial Dictatorship.} In this mechanism, bidders are ordered
(randomly or by priority), and each in turn selects from the remaining options.
Serial dictatorship is strategy-proof under its standard model, in which each
agent directly chooses its most preferred available alternative. An
item-by-item adaptation is not equivalent when bidders have complementarities
over slot bundles, because a sequence of individual picks need not represent
preferences over complete bundles. The administrative FAA priority lottery is
also not a direct-revelation auction and should not be assigned the same
incentive properties without specifying its choice protocol.

\paragraph{Our Mechanism (Learned Auction).} Our approach uses a neural
mechanism that outputs a fractional allocation rule over structured bundles and
an expected payment target. The allocation is resolved into a lottery using
Section~\ref{app:decomposition}, and realized payments use the conditional rule
described there. The mechanism is trained for low empirical regret and to
optimize a chosen revenue--welfare objective; these are approximate empirical
properties rather than exact guarantees.

\begin{table}[h]
\centering
\small
\begin{adjustbox}{max width=\textwidth}
\begin{tabular}{lcccc}
\toprule
\textbf{Mechanism} & \textbf{Efficiency} & \textbf{Incentive Compatible} & \textbf{Fairness} & \textbf{Revenue Support} \\
\midrule
Pure Lottery & No (in general) & Vacuous & Procedural only & No \\
Serial Dictatorship & No (in general) & Standard model only & Not guaranteed & No (unless extended) \\
\textbf{Our Mechanism} & Optimized & Low empirical regret & Soft target & Yes \\
\bottomrule
\end{tabular}
\end{adjustbox}
\caption{Comparison of mechanism properties.}
\end{table}

\paragraph{Interpretation.} The learned mechanism can represent rich bundle
preferences while optimizing revenue, welfare, and a soft fairness target.
Its fractional output is not itself a deployable schedule. We therefore use
the normalized discrete approximation and conditional proportional pricing in
Section~\ref{app:decomposition}. This implementation preserves feasibility and
reported-value ex-post IR, but it can change expected outcomes and incentive
properties; the resolved mechanism should consequently be evaluated on its own
revenue, welfare, fairness, and regret metrics.

\subsubsection{The 2011 Slot Auction}

While slot auctions have been proposed repeatedly — notably by the FAA in 2008–2009 — such proposals have met resistance from incumbent airlines and legal challenges, and have not been implemented. An instance of an actual slot auction at U.S. airports occurred in 2011, as a result of a DOT-imposed condition on the Delta–US Airways slot exchange at LGA and DCA. To preserve competition, the DOT required the divestiture of: 16 slot pairs at LGA, and 8 slot pairs at DCA. These were auctioned off via a sealed-bid, one-round auction, with eligibility restricted to carriers holding less than 5\% of the slots at the respective airport. The slots were sold in predefined bundles: Two bundles of 8 slot pairs at LGA, One bundle of 8 slot pairs at DCA. Winning bids were determined solely by price, and payments went directly to the divesting airlines. JetBlue and WestJet were the winning bidders. According to the GAO, JetBlue paid \$32 million for LGA slots and \$40 million for DCA slots, while WestJet paid \$17.6 million for the second LGA bundle.

The 2011 auction thus serves as a rare but valuable precedent. It confirms that sealed-bid, small-scale slot auctions are legally and operationally feasible under the right regulatory framework. This makes it a compelling test case for mechanism design approaches, particularly when applied to limited, high-value slot reallocations involving a small number of bidders and bundles.

\section{Greedy Discrete Approximation and Proportional Pricing}
\label{app:decomposition}

The neural mechanism produces a fractional allocation matrix
$Z \in [0,1]^{n \times |K|}$ satisfying the relaxed combinatorial
feasibility constraints. In the airport slot-allocation setting, however, the
implemented outcome must be a deterministic feasible schedule. This appendix
describes the resolution procedure used to map the fractional allocation to a
lottery over deterministic schedules, and the associated ex-post payment rule.

A key distinction is important. In simple assignment problems, fractional
feasibility can imply exact implementability as a lottery over deterministic
assignments. In the combinatorial setting considered here, overlapping bundles
can induce non-bihierarchical constraint structures, so fractional feasibility
alone does not guarantee that
$Z \in \operatorname{conv}(\mathcal{B})$, where $\mathcal{B}$ denotes the set of
feasible deterministic schedules. Accordingly, the procedure below should be
interpreted as a greedy discrete approximation rather than a guaranteed exact
convex decomposition.

\subsection{Greedy Discrete Approximation}

Let $\mathcal{B}$ be the set of all feasible deterministic allocation matrices
that satisfy the bidder-capacity and item-conflict constraints. Each
$D \in \mathcal{B}$ is binary, with $D_{iS}=1$ if bidder $i$ receives bundle
$S$, and $D_{iS}=0$ otherwise. We seek a sparse lottery over feasible schedules
whose expected allocation approximates the fractional allocation $Z$.

Finding the best sparse approximation over $\mathcal{B}$ is computationally
difficult, so we use a greedy iterative extraction procedure inspired by
Matching Pursuit \citep{matching_pursuit}. The algorithm repeatedly selects the
feasible deterministic schedule most aligned with the current residual and
extracts the largest nonnegative weight that preserves residual nonnegativity.

\begin{enumerate}
    \item \textbf{Initialize.} Set the residual matrix $R_0 = Z$, the iteration
    counter $t=0$, and the extracted lottery support
    $\Lambda = \emptyset$.

    \item \textbf{Iterative extraction.} While $\|R_t\|_1 > \epsilon$ and
    $t < T_{\max}$:
    \begin{enumerate}
        \item \emph{Select a feasible schedule.} Choose the feasible
        deterministic matrix with maximum inner product against the current
        residual:
        \[
        D_t \in \arg\max_{D \in \mathcal{B}} \langle R_t, D \rangle .
        \]
        This step identifies the feasible schedule that explains the largest
        remaining mass in the fractional allocation. Feasibility is downward
        closed, so any selected pair with zero residual can be deleted without
        changing the objective or violating feasibility. We use this reduced
        maximizer; if it is empty, the procedure terminates.

        \item \emph{Extract the maximum admissible weight.} Let
        \[
        \Omega_t = \{(i,S): D_{t,iS}=1\}.
        \]
        Set
        \[
        \gamma_t
        =
        \min_{(i,S)\in \Omega_t} R_{t,iS}.
        \]
        This is the largest value such that
        $R_t - \gamma_t D_t \geq 0$ elementwise. If
        $\gamma_t\leq\epsilon$, terminate; this prevents a
        zero-weight selection from stalling the procedure.

        \item \emph{Update the residual.} Set
        \[
        R_{t+1} = R_t - \gamma_t D_t,
        \]
        record the pair $(D_t,\gamma_t)$ in $\Lambda$, and increment $t$.
    \end{enumerate}

    \item \textbf{Construct the discrete approximation.} After $T$ iterations,
    define
    \[
    A_T = \sum_{t=0}^{T-1} \gamma_t D_t,
    \qquad
    R_T = Z - A_T.
    \]
    The lottery support consists of the extracted feasible schedules
    $\{D_t\}_{t=0}^{T-1}$ with extraction weights
    $\{\gamma_t\}_{t=0}^{T-1}$.
\end{enumerate}

The procedure guarantees residual nonnegativity and monotone residual reduction.
Because $\gamma_t$ is chosen as the minimum residual over the selected
support, we have $R_{t+1}\geq 0$ elementwise. Moreover,
\[
\|R_{t+1}\|_1
=
\|R_t\|_1 - \gamma_t \|D_t\|_1
\leq
\|R_t\|_1,
\]
with strict decrease whenever $\gamma_t>0$ and $D_t$ contains at least one
assigned bidder-bundle pair.

The extraction is exact as an implementation only if it exhausts the residual
and its weights form a subprobability distribution (the null schedule receives
any remaining mass). In general, including when the fractional point lies
outside $\operatorname{conv}(\mathcal{B})$, the raw extraction gives
\[
Z = A_T + R_T
=
\sum_{t=0}^{T-1} \gamma_t D_t + R_T.
\]
The raw extraction weights need not sum to one. To obtain a valid lottery,
define
\[
c_T=\max\left\{1,\sum_{t=0}^{T-1}\gamma_t\right\},
\qquad
\pi_t=\frac{\gamma_t}{c_T},
\qquad
\pi_{\emptyset}=1-\sum_{t=0}^{T-1}\pi_t.
\]
The null schedule $D^{\emptyset}=0$ receives probability
$\pi_{\emptyset}$. Hence the expected implemented allocation is
\[
\widehat Z_T
=\sum_{t=0}^{T-1}\pi_tD_t+\pi_{\emptyset}D^{\emptyset}
=\frac{A_T}{c_T}.
\]
Let $E_T=Z-\widehat Z_T$. For any valuation profile satisfying
$0\leq v_{iS}\leq v_{\max}$, H\"older's inequality gives
\[
\left|\langle v,Z\rangle-\langle v,\widehat Z_T\rangle\right|
=|\langle v,E_T\rangle|
\leq v_{\max}\|E_T\|_1.
\]
Thus $R_T$ measures the remaining mass after raw extraction, while $E_T$
measures the economic error of the normalized implementable lottery. When
$\sum_t\gamma_t\leq1$, normalization leaves the extracted weights unchanged
and $E_T=R_T$.

\subsection{Empirical Decomposition Audit}
\label{app:decomposition_audit}

We audit the preceding approximation on all 20{,}000 held-out airport profiles
using an independently retrained seed-0 CAGraph checkpoint. The rerun uses the
checked-in valuation file and airport configuration, reserves profiles
80{,}000--99{,}999 for evaluation, enumerates all 65 feasible schedules
(including the null schedule) in the selection oracle, sets
$\epsilon=10^{-10}$, and permits at most 10{,}000 extraction iterations. This
checkpoint is used only for the structural decomposition audit; the economic
estimates in the main airport tables remain the authoritative reported results.

\begin{table}[h]
\centering
\small
\setlength{\tabcolsep}{5pt}
\begin{tabular}{lrrrr}
\toprule
Metric & Mean & Median & 95th pct. & Maximum \\
\midrule
$\sum_t\gamma_t$ & 0.6883 & 0.6916 & 1.0000 & 1.0000 \\
Support size & 15.5 & 15.5 & 28 & 28 \\
$\lVert E_T\rVert_1$ & $3.81\!\times\!10^{-15}$
& $2.60\!\times\!10^{-15}$ & $9.18\!\times\!10^{-15}$
& $1.25\!\times\!10^{-14}$ \\
$v_{\max}\lVert E_T\rVert_1$ (dollars)
& $4.56\!\times\!10^{-9}$ & $3.11\!\times\!10^{-9}$
& $1.10\!\times\!10^{-8}$ & $1.50\!\times\!10^{-8}$ \\
\bottomrule
\end{tabular}
\caption{Discrete-decomposition audit on 20{,}000 held-out airport profiles.
The normalization factor was $c_T=1$ for every profile; hence $E_T=R_T$ and
the reported errors are at numerical precision.}
\label{tab:decomposition_audit}
\end{table}

No profile had $\sum_t\gamma_t>1$, so normalization changed neither the
extracted weights nor the expected allocation. The maximum absolute realized
welfare difference $|\langle v,E_T\rangle|$ was
$1.34\times10^{-8}$ dollars, below its corresponding H\"older certificate.
All extracted columns satisfied the bidder and item-capacity constraints. This
audit establishes numerical implementability for the evaluated allocations,
not universal implementability for arbitrary fractional points. It also does
not constitute a post-resolution incentive audit, which would require aligned
truthful and optimized-misreport outputs passed through the complete resolution
and conditional-pricing pipeline.

\subsection{Proportional Ex-Post Pricing}

The payment network outputs an expected payment $p_i$ for each bidder. Charging
$p_i$ as a non-refundable entrance fee would be inappropriate in a lottery
implementation, because a bidder could lose the lottery and still be charged.
This would violate ex-post individual rationality.

We therefore use a conditional proportional pricing rule. Let
$\widehat{Z}_T$ denote the expected allocation of the normalized lottery and
let $b_{iS}$ be bidder $i$'s reported value for bundle $S$. Define the expected
reported value delivered to bidder $i$ as
\[
\widehat{B}_i
=
\sum_{S\in K} \widehat{Z}_{T,iS} b_{iS}.
\]
For bidders with $\widehat{B}_i>0$, define the proportional charge rate
\[
\alpha_i
=
\min\left\{1,\frac{p_i}{\widehat{B}_i}\right\}.
\]
For bidders with $\widehat{B}_i=0$, set $\alpha_i=0$.

After a deterministic schedule is sampled, bidder $i$ is charged only if it
receives a bundle. If bidder $i$ receives bundle $S$, its ex-post payment is
\[
p_i^{\mathrm{ex post}}(S)
=
\alpha_i b_{iS}.
\]
If bidder $i$ receives no bundle, then
\[
p_i^{\mathrm{ex post}}(\emptyset)=0.
\]

This rule guarantees ex-post individual rationality relative to the submitted
values because $\alpha_i \leq 1$, so a winning bidder never pays more than its
reported value:
\[
b_{iS} - p_i^{\mathrm{ex post}}(S)
=
(1-\alpha_i)b_{iS}
\geq 0.
\]
Losers pay zero. Under truthful reporting, this is also ex-post individual
rationality with respect to true value. The rule is deployable because it uses
reported bids rather than unobserved true valuations.

When $p_i \leq \widehat{B}_i$, the expected ex-post payment exactly matches the
network's target payment:
\[
\mathbb{E}\left[p_i^{\mathrm{ex post}}\right]
=
\sum_{S\in K} \widehat{Z}_{T,iS}\alpha_i b_{iS}
=
\alpha_i \widehat{B}_i
=
p_i.
\]
When $p_i > \widehat{B}_i$, the cap $\alpha_i \leq 1$ binds, and expected
revenue is limited by the bidder's expected reported value. This case identifies
a discrepancy between the payment target and the implementable ex-post lottery,
and is included in the empirical discretization audit.

The decomposition and conditional payment transformation change the outcome
rule optimized during neural training. Consequently, a regret guarantee for the
fractional mechanism does not automatically transfer to the resolved mechanism;
post-resolution regret must be evaluated separately whenever incentive claims
are made for the deployed lottery.

\paragraph{Numerical example.}
Consider a truthfully reporting airline with a $50\%$ chance of winning a slot pair valued at
$\$40{,}000$. Its expected realized value is
\[
0.5 \times \$40{,}000 = \$20{,}000.
\]
If the payment network targets an expected payment of $\$10{,}000$, the
proportional charge rate is
\[
\alpha = \frac{10{,}000}{20{,}000}=0.5.
\]
If the airline wins the slot pair, it pays
\[
0.5 \times \$40{,}000 = \$20{,}000.
\]
If it loses, it pays $\$0$. The expected payment is therefore
\[
0.5 \times \$20{,}000 + 0.5 \times \$0
=
\$10{,}000,
\]
which recovers the target expected payment without charging any upfront entrance
fee and while preserving ex-post individual rationality.

\subsection{Optional Full-Assignment Variant}

The normalized lottery above may assign positive probability
$\pi_{\emptyset}$ to the null schedule. If an application requires a non-null
schedule whenever at least one is available, one may condition on the event
that the sampled schedule is non-null. The conditional probabilities are
\[
\pi_t^{+}=\frac{\pi_t}{1-\pi_{\emptyset}},
\qquad t=0,\ldots,T-1,
\]
provided $\pi_{\emptyset}<1$. Every realized schedule remains feasible, and
the proportional payment cap continues to give reported-value ex-post IR.
However, conditioning changes the expected allocation and generally changes
expected revenue, welfare, and regret. It therefore does not preserve the
fractional mechanism or its incentive properties and must be reported and
audited as a separate implementation variant.

\section{Computational Complexity}
\label{app:complexity}

Let $n$ denote the number of bidders, $m$ the number of items, and
$k=|K|$ the number of candidate bundles (equal to $2^m-1$ when every
nonempty bundle is enumerated). In the fully expressive
combinatorial representation, each bidder can submit a value for each
bundle, so the bid tensor contains $O(nk)$ entries. This exponential
growth in $k$ is the dominant scalability bottleneck of any fully
enumerative combinatorial auction method, including ours.

For CANet, the input valuation vector is flattened and passed through
$L$ fully connected layers of hidden dimension $H$. Ignoring batch size,
the forward-pass complexity is
\[
O(nkH + LH^2),
\]
where the first term corresponds to the input and output projections and
the second term corresponds to the hidden layers.

For CAFormer, the complexity depends on the number of tokens used to
represent bidders, items, and bundles. With $O(n+k)$ tokens and hidden
dimension $H$, each self-attention layer has worst-case complexity
\[
O((n+k)^2H + (n+k)H^2).
\]
Thus, CAFormer can be expressive but becomes increasingly expensive as
the number of feasible bundles grows.

For CAGraph, we construct a conflict graph whose nodes correspond to
bidder-bundle pairs. Hence, the graph has
\[
|V| = O(nk)
\]
nodes. Edges connect mutually incompatible allocations, including
allocations involving overlapping bundles or the same bidder. A graph
attention layer has complexity
\[
O(|E|H),
\]
where $|E|$ is the number of conflict edges. In the worst case,
$|E|=O(n^2k^2)$, but in practical slot-allocation instances the graph is
substantially sparser because edges are induced only by actual bundle
conflicts. Moreover, the airport application uses a filtered set of
operationally valid bundles rather than all $2^m-1$ possible bundles.

The Capacity Normalization layer adds only polynomial overhead relative
to the enumerated bundle set. Its operations consist primarily of
normalization and masking over bidder-bundle and item-bundle tensors,
which scale as
\[
O(nk + mk).
\]
Therefore, the constraint layer does not change the fundamental
exponential dependence on $m$; rather, it ensures that the network output
satisfies the combinatorial feasibility constraints by construction.

For the Matching Pursuit-inspired discrete decomposition, let $T$ denote
the number of decomposition iterations and let $\mathcal{B}$ denote the
set of deterministic feasible allocations considered by the oracle. At
each iteration, the algorithm selects
\[
D^* = \arg\max_{D \in \mathcal{B}} \langle R, D \rangle,
\]
computes the largest feasible extraction weight $\gamma$, and updates
the residual. If the selection step is solved by explicit enumeration
over $\mathcal{B}$, each iteration requires
\[
O(|\mathcal{B}| \cdot nk)
\]
operations to evaluate the inner products, yielding total complexity
\[
O(T |\mathcal{B}| nk).
\]
However, explicit enumeration is generally impractical because
$|\mathcal{B}|$ can be exponential in the number of items and bidders.
The selection step can be formulated as a weighted winner-determination
problem over the conflict graph, where node weights are the current residual
entries. For the small airport instance, explicit enumeration of the feasible
schedules is also tractable. In either implementation, the worst-case
complexity remains exponential, as winner determination in combinatorial
auctions is NP-hard, but the practical cost is controlled by the number of
valid bundles retained after operational filtering.

The residual update and extraction step are inexpensive relative to the
selection oracle. Given $D^*$, computing $\gamma$ and updating $R$
require $O(nk)$ operations. Therefore, the dominant cost of the
decomposition is the repeated solution of the deterministic allocation
oracle, not the residual update itself. In the airport-slot case study,
this step is tractable because the number of feasible bundles is small
after filtering, and the decomposition is applied only after the neural
mechanism has been trained.

Overall, the dominant computational bottlenecks are the exponential
growth of the bundle set and, for discrete implementation, the repeated
winner-determination oracle used in the Matching Pursuit decomposition.
The neural feasibility layer itself adds only polynomial overhead in the
number of enumerated bidder-bundle variables. The proposed framework is
therefore most appropriate for targeted combinatorial allocation
problems with a moderate number of valid bundles, such as regulatory
slot divestitures, spectrum packages, or infrastructure-capacity
assignments.

\section{Empirical Regret Audit}
\label{app:regret_audit}

The audit reports the best deviation found by the configured gradient-based
misreport search on held-out profiles. It is a lower bound on true ex-post
regret and should not be read as a proof of DSIC. The dollar column applies the
same inverse normalization used to return airport valuations and payments to
monetary units. Because the fairness term is implemented as a soft target, the
labels below refer to target Jain indices rather than certified lower bounds.
Revenue gains use the rounded \$139{,}000 VCG benchmark reported in the main
frontier table.

\begin{table}[h]
\centering
\small
\caption{Empirical regret audit for the revenue--welfare frontier. Revenue gains are measured relative to VCG.}
\label{tab:appendix_regret_revenue_welfare}
\begin{adjustbox}{max width=\textwidth}
\begin{tabular}{lrrrr}
\toprule
Mechanism & Revenue gain vs. VCG (\$) & Normalized mean regret & Dollar mean regret (\$) & Regret / revenue gain \\
\midrule
$\lambda=0$    & 11{,}204  & 0.0033037 & 19.6025 & 0.001750 \\
$\lambda=0.25$ & 229{,}957 & 0.0000695 & 0.4121  & 0.000002 \\
$\lambda=0.5$  & 237{,}201 & 0.0000175 & 0.1040  & 0.000000 \\
$\lambda=0.75$ & 237{,}870 & 0.0002827 & 1.6774  & 0.000007 \\
$\lambda=1$    & 252{,}409 & 0.0000000 & 0.0000  & 0.000000 \\
\bottomrule
\end{tabular}
\end{adjustbox}
\end{table}

\begin{table}[h]
\centering
\small
\caption{Empirical regret audit for the welfare--fairness frontier. Revenue gains are measured relative to VCG.}
\label{tab:appendix_regret_welfare_fairness}
\begin{adjustbox}{max width=\textwidth}
\begin{tabular}{lrrrr}
\toprule
Mechanism & Revenue gain vs. VCG (\$) & Normalized mean regret & Dollar mean regret (\$) & Regret / revenue gain \\
\midrule
Target $J=0.25$  & 11{,}366  & 0.0093867 & 55.6966 & 0.004900 \\
Target $J=0.50$  & 252{,}639 & 0.0014449 & 8.5734  & 0.000034 \\
Target $J=0.75$  & 235{,}420 & 0.0011410 & 6.7703  & 0.000029 \\
Target $J=0.875$ & 249{,}067 & 0.0001966 & 1.1665  & 0.000005 \\
\bottomrule
\end{tabular}
\end{adjustbox}
\end{table}

\begin{table}[h]
\centering
\small
\caption{Empirical regret audit for the revenue--fairness frontier. Revenue gains are measured relative to VCG.}
\label{tab:appendix_regret_revenue_fairness}
\begin{adjustbox}{max width=\textwidth}
\begin{tabular}{lrrrr}
\toprule
Mechanism & Revenue gain vs. VCG (\$) & Normalized mean regret & Dollar mean regret (\$) & Regret / revenue gain \\
\midrule
Target $J=0.25$  & 257{,}103 & 0.0000007 & 0.0041  & 0.000000 \\
Target $J=0.50$  & 243{,}778 & 0.0000015 & 0.0088  & 0.000000 \\
Target $J=0.75$  & 241{,}961 & 0.0023313 & 13.8328 & 0.000057 \\
Target $J=0.875$ & 240{,}054 & 0.0000121 & 0.0720  & 0.000000 \\
\bottomrule
\end{tabular}
\end{adjustbox}
\end{table}

\section{Scaling to Real-World Problems}
\label{app:scaling}

To transition from stylized divestitures to national-scale airport auctions, we identify three architectural requirements for a production-ready system.

\subsection{Non-Homogeneous Resource Capacity}
In practice, airport slots are organized into temporal blocks $B_1, \dots, B_K$ with varying throughput capacities $C_k$ (e.g., lower capacity during noise-abatement or maintenance windows). We propose extending \textbf{CAGraph} by defining a resource-weighted conflict graph where edges represent the marginal consumption of a block's total capacity. The allocation $z$ must satisfy:
\begin{equation}
    \sum_{i \in \mathcal{N}} \sum_{s \in \mathcal{S}} z_{i,s} \cdot \omega_{s,k} \le C_k, \quad \forall k \in \{1, \dots, K\}
\end{equation}
where $\omega_{s,k} \in \{0,1\}$ denotes if bundle $s$ utilizes a slot in block $k$. This allows for arbitrary capacities (e.g., 2 slots in one block, 4 in another). Additionally, we suggest changing the standard Softmax that forces the sum of item to exact 1, to a normalization layer that caps the sum at $C_k$.

\subsection{Hierarchical Decomposition for $N$ Bidders}
As the number of bidders $N$ and blocks $K$ increase, the global winner determination problem faces combinatorial explosion. We propose a hierarchical structure:
\begin{itemize}
    \item \textbf{Spectral Partitioning:} Use graph clustering to decompose the global schedule into $K$ nearly-independent sub-problems.
    \item \textbf{Parallel Neural Clearing:} Deploy independent \textbf{CAGraph} instances to clear sub-blocks in parallel. The resulting runtime and approximation quality would require empirical evaluation; no linear-time or millisecond guarantee is claimed.
    \item \textbf{Global Coordination:} A high-level transformer layer manages coupling constraints, such as ensuring a carrier's arrival in an early block has a corresponding feasible departure in a subsequent block.
\end{itemize}

\subsection{Scalable Candidate Generation}
To handle high-dimensional bidder sets, future work could use a candidate-generation
submodule. Such a layer would identify a reduced active set of bundles for each
bidder and pass it to the neural mechanism, avoiding enumeration of the full
power set. Because report-dependent candidate generation can alter allocation
monotonicity and profitable deviations, incentive compatibility would need to
be re-audited for the combined system rather than assumed to be preserved.


\end{document}